\documentclass[fleqn,usenatbib]{mnras}
\usepackage{newtxtext,newtxmath}

\usepackage[T1]{fontenc}

\usepackage{graphicx}	% Including figure files
\usepackage{amsmath}	% Advanced maths commands
\usepackage{xcolor}
\usepackage{ulem}
\newcommand{\wt}{\omega(\theta)}

\newcommand{\cgg}{C_\mathrm{gg}}
\newcommand{\cgk}{C_\mathrm{g\kappa}}

\newcommand{\pgg}{P_\mathrm{gg}}
\newcommand{\pgk}{P_\mathrm{g\kappa}}

\newcommand{\qg}{q_{\mathrm{g}}}
\newcommand{\qk}{q_{\mathrm{\kappa}}}

\newcommand\rossi[1]{\textcolor{black}{#1}}

\newcommand\rossiiii[1]{\textcolor{black}{#1}}

\newcommand{\be}{\begin{equation}}
\newcommand{\ee}{\end{equation}}
\newcommand{\ba}{\begin{eqnarray}}
\newcommand{\ea}{\end{eqnarray}}

\newcommand{\mx}{\mbox}
\newcommand{\bm}{\boldmath}
\newcommand{\p}{\mbox{\boldmath $p$}}

\newcommand{\muu}{\mbox{\boldmath $\mu$}}
\newcommand{\Modcomp}{\mbox{\boldmath $\mathcal{M}$ } }

\def\btheta{{\mx {\bm $\theta$}}}
\def\bphi{{\mx {\bm $\phi$}}}

\def\deg2{\rm deg^2}
\def\arcmin2{\rm arcmin^2}

\def\nn{{\nonumber}}

\def\s8{{\sigma_8}}

\def\kpc{\, h^{-1}{\rm kpc}}

\title[Data compression for DESI+Lensing]{A data compression and optimal galaxy weights scheme for Dark Energy Spectroscopic Instrument and weak lensing datasets}

\author[Ruggeri et al.]{
\parbox{\textwidth}{
\Large
Rossana~Ruggeri,$^{1,2}$
Chris~Blake,$^{2}$
Joseph~DeRose,$^{3}$
C.~Garcia-Quintero,$^{4}$
B.~Hadzhiyska,$^{5}$
M.~Ishak,$^{4}$
N.~Jeffrey,$^{6}$
S.~Joudaki,$^{7}$
Alex~Krolewski,$^{7,8,9}$
J.~U.~Lange,$^{10,11}$
A.~Leauthaud,$^{12}$
A.~Porredon,$^{13,14}$
G.~Rossi,$^{15}$
C.~Saulder,$^{16}$
E.~Xhakaj,$^{12}$
D.~Brooks,$^{6}$
G.~Dhungana,$^{17}$
A.~de la Macorra,$^{18}$
P.~Doel,$^{6}$
S.~Gontcho A Gontcho,$^{3,19}$
A.~Kremin,$^{3}$
M.~Landriau,$^{3}$
R.~Miquel,$^{20,21}$
C.~Poppett,$^{3,22,23}$
F.~Prada,$^{24}$
and Gregory~Tarl\'{e}$^{25}$
\begin{center} (DESI Collaboration) \end{center}
}
\vspace{0.4cm}
\\
\parbox{\textwidth}{
$^{1}$ School of Mathematics and Physics, University of Queensland, 4072, Australia\\
$^{2}$ Centre for Astrophysics \& Supercomputing, Swinburne University of Technology, P.O. Box 218, Hawthorn, VIC 3122, Australia\\
$^{3}$ Lawrence Berkeley National Laboratory, 1 Cyclotron Road, Berkeley, CA 94720, USA\\
$^{4}$ The University of Texas at Dallas, 800 W. Campbell Rd., Richardson, TX 75080, USA\\
$^{5}$ Center for Astrophysics $|$ Harvard \& Smithsonian, 60 Garden Street, Cambridge, MA 02138, USA\\
$^{6}$ Department of Physics \& Astronomy, University College London, Gower Street, London, WC1E 6BT, UK\\
$^{7}$ Department of Physics and Astronomy, University of Waterloo, 200 University Ave W, Waterloo, ON N2L 3G1, Canada\\
$^{8}$ Perimeter Institute for Theoretical Physics, 31 Caroline St. North, Waterloo, ON N2L 2Y5, Canada\\
$^{9}$ Waterloo Centre for Astrophysics, University of Waterloo, 200 University Ave W, Waterloo, ON N2L 3G1, Canada\\
$^{10}$ Kavli Institute for Particle Astrophysics and Cosmology, Stanford University, Menlo Park, CA 94305, USA\\
$^{11}$ Physics Department, Stanford University, Stanford, CA 93405, USA\\
$^{12}$ Department of Astronomy and Astrophysics, University of California, Santa Cruz, 1156 High Street, Santa Cruz, CA 95065, USA\\
$^{13}$ Center for Cosmology and Astro-Particle Physics, The Ohio State University, Columbus, OH 43210, USA\\
$^{14}$ Department of Physics, The Ohio State University, Columbus, OH 43210, USA\\
$^{15}$ Department of Physics and Astronomy, Sejong University, Seoul, 143-747, Korea\\
$^{16}$ Korea Astronomy and Space Science Institute, 776, Daedeokdae-ro, Yuseong-gu, Daejeon 34055, Republic of Korea\\
$^{17}$ Department of Physics, Southern Methodist University, 3215 Daniel Avenue, Dallas, TX 75275, USA\\
$^{18}$ Instituto de F\'{\i}sica, Universidad Nacional Aut\'{o}noma de M\'{e}xico,  Cd. de M\'{e}xico  C.P. 04510,  M\'{e}xico\\
$^{19}$ Department of Physics and Astronomy, University of Rochester, 500 Joseph C. Wilson Boulevard, Rochester, NY 14627, USA\\
$^{20}$ Instituci\'{o} Catalana de Recerca i Estudis Avan\c{c}ats, Passeig de Llu\'{\i}s Companys, 23, 08010 Barcelona, Spain\\
$^{21}$ Institut de F\'{i}sica d’Altes Energies (IFAE), The Barcelona Institute of Science and Technology, Campus UAB, 08193 Bellaterra Barcelona, Spain\\
$^{22}$ Space Sciences Laboratory, University of California, Berkeley, 7 Gauss Way, Berkeley, CA  94720, USA\\
$^{23}$ University of California, Berkeley, 110 Sproul Hall \#5800 Berkeley, CA 94720, USA\\
$^{24}$ Instituto de Astrofisica de Andaluc\'{i}a, Glorieta de la Astronom\'{i}a, s/n, E-18008 Granada, Spain\\
$^{25}$ University of Michigan, Ann Arbor, MI 48109, USA\\
}
}

\date{Accepted XXX. Received YYY; in original form ZZZ}

\pubyear{2015}

\begin{document}
\label{firstpage}
\pagerange{\pageref{firstpage}--\pageref{lastpage}}
\maketitle

\begin{abstract}
   Combining different observational probes, such as galaxy clustering and weak lensing, is a promising technique for unveiling the physics of the Universe with upcoming dark energy experiments.  The galaxy redshift sample from the Dark Energy Spectroscopic Instrument (DESI) will have a significant overlap with major ongoing imaging surveys specifically designed for weak lensing measurements: the Kilo-Degree Survey (KiDS), the Dark Energy Survey (DES) and the Hyper Suprime-Cam (HSC) survey.  In this work we analyse simulated redshift and lensing catalogues to establish a new strategy for combining high-quality cosmological imaging and spectroscopic data, in view of the first-year data assembly analysis of DESI.  In a test case fitting for a reduced parameter set, we employ an optimal data compression scheme able to identify those aspects of the data that are most sensitive to the cosmological information, and amplify them with respect to other aspects of the data.  We find this optimal compression approach is able to preserve all the information related to the growth of structure; we also extend this scheme to derive weights to be applied to individual galaxies, and show that these produce near-optimal results.
\end{abstract}


%%%%%%%%%%%%%%%%%%%%%%%%%%%%%%%%%%%%%%%%%%%%%%%%%%

%%%%%%%%%%%%%%%%% BODY OF PAPER %%%%%%%%%%%%%%%%%%

\section{Introduction}

Over the next five years, the Dark Energy Spectroscopic Instrument (DESI) will map the position of $\gtrsim 30$ million galaxies, reconstructing the cosmic history from the nearby universe to a distance of 11 billion light-years \citep{2016arXiv161100036D}. This unprecedented volume of data, 10-15 times the size of current samples, will provide insights to fundamental questions unanswered by the current standard cosmological model.

Some of the key science questions that DESI aims to address will benefit from combining DESI data with other cosmological probes.  In this paper we focus on the potential of combining DESI and weak gravitational lensing data.  DESI has a significant overlap with ongoing deep imaging surveys specifically designed for weak lensing measurements: the Kilo-Degree Survey (KiDS) \citep{2013ExA....35...25D}; the Dark Energy Survey (DES) \citep{2016MNRAS.460.1270D} and the Hyper Suprime-Cam (HSC) survey \citep{2018PASJ...70S..25M}.  The DESI Year 1 dataset is expected to have at least $1{,}000$ deg$^2$ overlap with these lensing surveys.  Combining the cosmological information contained in DESI large-scale structure and external lensing datasets will allow for broader exploration of new theories of gravity, by measuring how modifications to the current theory of gravity would affect light and matter simultaneously \citep[e.g.,][]{2020JCAP...12..018G, 2018MNRAS.474.4894J, 2010PhRvD..81l3508D}.  A joint analysis will improve measurements of the parameters of interest, decrease the degeneracies, and help mitigate systematic errors not controlled in an individual analysis.

Whilst combined analysis of these new datasets will push our understanding of the Universe to the next level, we wish to emphasise two key, linked challenges to be solved.  The first challenge is how to optimally weight the data to achieve a given scientific outcome (for example, how to weight a DESI galaxy contributing to both galaxy-galaxy lensing and clustering measurements).  The second challenge is how to make tractable the big-data volumes produced by DESI using techniques such as data compression.  In this context, an important advantage of data compression is mitigating the challenge of the combined-probe covariance.  After separating all the tomographic and separation bins, a combined analysis of DESI and weak lensing data could utilise many hundreds or even thousands of data points with significant inter-correlations.  If the covariance is estimated using a standard approach of performing similar measurements on mock catalogues, an intractable number of mocks may be required to maintain a low level of noise \citep{1967Kaufman, 2007A&A...464..399H, 2022MNRAS.510.3207P, 2020MNRAS.498.3744R}.  Whilst analytical covariance estimates are also possible, these may not accurately incorporate details such as the survey footprint or other effects such as non-linearities or fibre collisions.

Previous studies have developed optimal weighting schemes for data compression with a focus on measuring the growth rate of structure \citep{2017MNRAS.464.2698R, 2019MNRAS.484.4100R, 2019MNRAS.483.3878R, 2019MNRAS.482.3497Z}, angular diameter distance \citep{2018MNRAS.480.1096Z},  primordial non Gaussianity \citep{2019JCAP...09..010C} and cosmic shear \citep{2019OJAp....2E..11B}. \cite{2020arXiv200506551M} present an application of the MOPED algorithm \citep{2000MNRAS.317..965H} to weak lensing measurements. 
These studies explored optimal weighting for measurements with individual probes, demonstrating how an optimal weighting scheme applied to a dataset gives unbiased results and is efficient in decreasing the computational costs. 
\rossi{Our compression method is similar to MOPED, but our applications are different to those highlighted by previous MOPED papers.  
We apply this data-compression to galaxy-galaxy lensing while \cite{2020arXiv200506551M} apply data-compression to cosmic shear. 
Further we here derive an optimal weighting scheme to be applied directly to galaxies, and derive a new weighted estimator that allows us to do that.  }

% Further we discuss here 
% To adjust: Emphasise point of difference with MOPED paper. Galaxy weights is one point. I would also add an extra sentence in paragraph 4 and 5. We apply this data-compression to these g-g lensing and clustering (different combination of data).  Other paper: apply only to cosmic shear.

% The application of a data-compression algorithm such as MOPED \citep{2000MNRAS.317..965H} to weak lensing measurements, which compress in scale and redshifts different measurements of the correlation functions. }

In \cite{2020MNRAS.498.2948R} we developed and tested an efficient way to combine information from galaxies at different epochs in the evolution of the Universe using lensing and clustering statistics in Fourier space, discussing the advantages of such an approach with respect to the binning of the galaxies adopted by past analyses and testing the optimal weighting technique on Gaussian realisations. 

In this work we extend the study of \cite{2020MNRAS.498.2948R} by applying the data-compression methodology to configuration-space statistics of clustering and galaxy-galaxy lensing measured from realistic mock simulations modelling the first year of DESI data.  For the purposes of this study we utilise a small parameter set, fixing the majority of the cosmological parameters and focusing on fitting for the overall normalisation of the power spectrum $\sigma_8$ and the galaxy bias parameters as a test case, although these techniques can be readily extended to a wider parameter set.

We also derive a weighting scheme that can be applied directly to galaxies, instead of to the power spectra or correlation function measurements.  In a similar approach to ``FKP weighting'' \citep{1994ApJ...426...23F}, by fixing an effective scale we generate weights which can be applied to individual galaxies, provided that the weights used are a smooth function on the range of scales considered.  We hence calculate optimal weights for the lenses for a combined-probe analysis for the first time, extending FKP weights to include other probes.  If the redshift bins are thin enough that the evolution in redshift within each bin is negligible, and large enough that the lensing signal can be measured,  galaxy-weighting and correlation function weighting give identical results. However, the galaxy weighting approach is independent of the width of the redshift bins considered, provided that enough freedom is allowed in describing all parameters in the model as a function of redshift.

The paper is organized as follows: in Sec. \ref{sec:model} we briefly review the models of the combined-probes statistics and their covariance that we use in our analysis.  In Sec. \ref{sec:datacompression} we derive the optimal weighting scheme for configuration-space statistics and the weights to be applied to individual objects with respect to a set of parameters. Sec. \ref{sec:measurements} describes the mock data used for the analysis and the estimator employed to measure the angular correlation functions in both lensing and clustering.  In Sec. \ref{sec:results} we present the test results from data-compression, and compare them with an uncompressed analysis.  We conclude in Sec. \ref{sec:conc}.

\section{Models and covariance}
\label{sec:model}

In this section we summarise our theoretical model for the combined-probe statistics we use in this study: the average tangential shear $\gamma_t(\theta)$ around lens galaxies and their angular clustering $\omega(\theta)$, along with the covariance of these statistics.

\subsection{Angular power spectra for combined probes}
\label{subsec:model}

We define the angular (cross-)power spectra between two fields $\delta_\mathrm{a}$, $\delta_\mathrm{b}$ measured in redshift bins $i,j$, as a function of the projected Fourier mode, $\ell$, as
\begin{equation}
 C_{\mathrm{ab}}^{ij}(\ell) = \int d\chi \frac{q_\mathrm{a}^i(\chi)q_\mathrm{b}^j(\chi)}{\chi^2}P_{\mathrm{ab}}(\ell/\chi,z(\chi)),
\label{eq:cab}
\end{equation}
where $P_{\mathrm{ab}}(k,z)$ is the 3D (cross-)power spectrum of the fields at redshift $z$ and wavenumber $k$, and $\chi(z)$ is the comoving distance \citep{1992ApJ...388..272K,PhysRevD.70.043009}.   We note that Eq. \ref{eq:cab} is derived assuming the Limber and flat-sky approximations  \citep{2017JCAP...05..014L, 2017MNRAS.472.2126K}.

The weight function  $q_\mathrm{a,b}(\chi)$ depends on the field considered. For the galaxy density field $\delta_\mathrm{g}$, $\qg(\chi)$ is defined as 
\begin{equation}
\label{eq:qg}
q_\mathrm{g}^i(\chi) =\frac{n^i_{\mathrm{lens}}(z)  }{\bar{n}_{\mathrm{lens}}^i}\frac{dz}{d\chi}\,,
\end{equation}
where $n^i_{\mathrm{lens}}(z)$ is the lens redshift distribution of sample $i$, with $z$ the redshift corresponding to $\chi$, and $\bar{n}_{\mathrm{lens}}^i$  is the average lens density.  For the convergence field $\delta_\kappa$,  $\qk(\chi)$ is  determined by the lensing efficiency,
 \begin{equation}
\label{eq:qkappa}
\qk^{i}(\chi) = \frac{3 H_0^2 \Omega_m }{2 c^2}\frac{\chi}{a(\chi)}\int_\chi^{\chi_{\mathrm{max}}} \, d \chi' \frac{n_{\mathrm{source}}^{i}(z)\,
%\left[ z(\chi') \right]
}{\bar{n}_{\mathrm{source}}^{i}}\frac{dz}{d\chi'} \frac{(\chi'-\chi)}{\chi'} \,,
\end{equation}
where $H_0$ and $\Omega_m$ denote the values of the present-day Hubble parameter and matter density, $c$ is the speed of light, $\chi_{\mathrm{max}}$ is the maximum comoving distance of the source distribution, $n^i_{\mathrm{source}}(z)$ and  $\bar{n}^i_{\mathrm{source}}$ are the source redshift distribution and average density of sources in sample $i$, and $a(\chi)$ is the cosmic scale factor.  We describe the specific source and lens configurations used in our study in Sec. \ref{sec:simulation} below.

\subsection{Covariance of angular power spectra}
The Gaussian covariance matrix between two angular power spectra  $ C_{\mathrm{ab}}^{ij} (\ell_1), C_{\mathrm{cd}}^{kl} (\ell_2)  $, for samples $(i,j,k,l)$, \rossi{in the absence of survey window effects}, is given by \cite{PhysRevD.70.043009} and  \cite{2017MNRAS.470.2100K},
\begin{equation}\begin{split}\label{eq:covpred}
\mathbf{C}  =& \frac{4 \pi \delta_{\ell_1 \ell_2}}{ \Omega_{\rm{s}} (2\ell_1+1) \Delta \ell_1}  \times\large[ \\
&\left(C_{\mathrm{ac}}^{ik}(\ell_1)+ \delta_{ik}\delta_{\mathrm{ac}} N_{\mathrm a}^i\right) \left(C_{\mathrm{bd}}^{jl}(\ell_2)+ \delta_{jl}\delta_{\mathrm{bd}}N_{\mathrm b}^j\right) \\
+&\left(C_{\mathrm{ad}}^{il}(\ell_1)+ \delta_{il}\delta_{\mathrm{ad}} N_{\mathrm a}^i\right) \left. \left(C_{\mathrm{bc}}^{jk}(\ell_2)+ \delta_{jk}\delta_{\mathrm{bc}} N_{\mathrm b}^j\right) \right]
%\,,    
\end{split}
\end{equation}
where $\Omega_{\rm{s}}$ is the angular area of the overlapping sample in steradians.  The noise terms are,
\begin{equation}\begin{split}
    N_{\mathrm{g}} &= 1/\bar{n}_{\mathrm{lens}},\\
    N_{\mathrm{\kappa}} &= \sigma_e^2 /\bar{n}_{\mathrm{source}},
\end{split}    
\end{equation}
where $\sigma_e$ is the shape noise.

\subsection{Configuration-space statistics}

In configuration-space measurements of galaxy-galaxy lensing, we work with the average tangential shear $\langle \gamma_t(\theta)\rangle $. This quantity defines the tangential shear of background galaxies at angular separation $\theta$ from a lens galaxy, and is related to the convergence $\kappa$ as, 
\begin{equation}
    \langle \gamma_t (\theta) \rangle = \langle \overline{\kappa} (< \theta) - \langle \kappa (\theta) \rangle ,
\end{equation}
where $ \overline{\kappa} (< \theta) $ is the integrated convergence within separation $\theta$. $\gamma_t$ can be described in terms of the $\cgk$ power spectrum via,
\begin{equation}
\label{eq:gammatamodel}
    \gamma_t (\theta) = \int \frac{d\ell \, \ell }{ 2 \pi} \, \cgk(\ell) \, J_2(\ell \theta) ,
\end{equation}
where $J_n$ denotes the Bessel function of the first kind.

We also define the angular correlation function $\omega(\theta)$ of the lens galaxies at separation $\theta$, which can be formulated as a function of the galaxy-galaxy angular power spectrum $\cgg$ as,
\begin{equation}
\label{eq:wtetamodel}
    \omega(\theta) = \int \frac{d\ell \, \ell }{ 2 \pi} \, \cgg(\ell) \, J_0(\ell \theta) .
\end{equation}
See e.g. \cite{2020A&A...642A.158B}.
\subsection{Covariance of configuration-space statistics}
\label{sec:cov}

The Gaussian covariance matrix \textbf{C}, between $\gamma_t(\theta)$ and $\wt$ for $N_\theta$ separation bins is composed of three main blocks: 
%$ \rm{Cov}\left(\gamma_t\right)(\theta_i, \theta_j) $,$\rm{Cov}\left(\gamma_t \omega\right) (\theta_i, \theta_j)$  $\langle \gamma_t (\theta_i ) \omega(\theta_j) \rangle $ and $\langle \omega (\theta_i ) \omega(\theta_j)  \rangle $. Where
 \ba
{\mathbf{C}}[ {\gamma}_t(\theta_i) , {\gamma}_t(\theta_j)] & = & 
\frac{1}{\Omega_s} \int_0^{\infty} \frac{d^2 l}{(2\pi)^2}\, 
{J}_2(l\theta_i) J_2(l\theta_j) \times \nn \\
& & \hspace{-1.6cm}\left\{ \mathcal{C}_\mathrm{g\kappa}^2(l) + \left[\mathcal{C}_\mathrm{\kappa
      \kappa}(l) +  N_{\mathrm{\kappa}}\right] 
  \left[\mathcal{C}_{\rm gg}(l) + N_{\mathrm{g}} \right]
 \right\},
\label{eq:cov_gamma_T}
\ea

 \ba
{\mathbf{C}}[ {\omega}(\theta_i) , {\omega}(\theta_j)] & = & 
\frac{2}{\Omega_s} \int_0^{\infty} \frac{d^2 l}{(2\pi)^2}\, 
{J}_0(l\theta_i) J_0(l\theta_j) \times \nn \\
& & \hspace{-1.6cm}\left\{ \left[\mathcal{C}_{\rm gg}(l) + N_{\mathrm{g}} \right]
  \left[\mathcal{C}_{\rm gg}(l) + N_{\mathrm{g}} \right]
 \right\},
\label{eq:cov_w}
\ea
\ba
{\mathbf{C}}[ \omega(\theta_i), { {\gamma}}_t(\theta_j)] 
\hspace{-0.2cm} & = & \hspace{-0.2cm} \frac{2}{\Omega_s}
\int \frac{d^2l}{(2\pi)^2} 
 {J}_0(l \theta_i) {J}_2(l \theta_j)  \nn\\ 
\hspace{-0.2cm} & \times & \hspace{-0.2cm} 
\mathcal{C}_\mathrm{g\kappa}(l)\left[\mathcal{C}_{\rm gg}(l) + N_{\mathrm{g}}\right].
\label{eq:cross_cov_BA}
\ea
where for each separation bin, we integrated Eqs. \ref{eq:cov_gamma_T}, \ref{eq:cov_w} and \ref{eq:cross_cov_BA} over the angular area of the bin
\citep{PhysRevD.70.043009}.
%I needed to average the correlation function model over bins, rather than evaluate at a single theta scale -- those averages use the same integrals as the covariance -- codes such as CosmoSIS do a bin averaging instead
%I was using the original halofit model in CAMB -- I should have been using the updated Takahashi et al. model -- pars.NonLinearModel.set_params(halofit_version='takahashi')

\subsection{Fiducial cosmology}
\label{subsec:fidcos}

We generated the non-linear matter power spectrum used in these model computations using the CAMB software \citep{Lewis:2002ah}, where we adopted the re-calibrated halofit model of \cite{2012ApJ...761..152T}.  We assumed a fiducial cosmological model consistent with the Buzzard simulations \citep{2019arXiv190102401D} introduced in Sec. \ref{sec:simulation}, with a matter density $\Omega_m = 0.286$, baryon density $\Omega_b = 0.047$, Hubble parameter $h =0.7$, amplitude of matter clustering $\sigma_8 = 0.82$ and spectral index $n_s = 0.96$.

To model the galaxy-galaxy and galaxy-convergence power spectra  $\pgg$ and $\pgk$, we assume a linear galaxy bias relation where $\pgg \propto b^2 \sigma_8^2$ and $\pgk \propto  b \sigma_8^2$.  Due to the selection of our mock DESI samples described in Sec. \ref{sec:simulation}, the galaxy bias evolves with redshift.  We describe this redshift-dependent relation using a Taylor expansion up to second order,
\begin{equation}
\label{eq:bias}
    b (z) =  b_{\mathrm{0}} + b_1 z + b_2 z^2 ,
    %
    %\frac{D(z_{\mathrm{piv}})}{D(z)},
\end{equation}
where $b_0$, $b_1$ and $b_2$ are free parameters.  We tested that a second-order model was sufficient to capture the galaxy bias evolution present in our mocks, and adding higher-order terms produced no significant change in our conclusions: \rossi{see Figure \ref{fig:bias_quad} and related discussion in Sec. \ref{subsec:compfit}}.
\rossiiii{ Note that the spectroscopic sample of DESI includes two different targets: Bright Galaxy Survey (BGS) $ 0.1 < z < 0.5 $ and  Luminous Red Galaxy Survey (LRGS)  $0.5<z<0.9$. For simplicity,  we adopt a unique bias-redshift relationship rather than a discontinuous function. This assumption is supported from tests on the mocks, where we observe a smooth  bias redshift relationship.   This fact is discussed further in Sec. \ref{subsec:uncompfit}}.
%1) Add comment on BGS -  LRGS -  Fig 3 shows a smooth transition in bias - we would see a big discontinuity at 0.5 - on the same trend. 
% where $D(z)$ is the linear growth rate and we selected $b_{\mathrm{piv}} = 2$ as the value of the galaxy bias at the pivot redshift $z_{\mathrm{piv}} = 0.45$.
% This relation is approximately correct for the clustering amplitude of magnitude-selected galaxy samples \citep{2001AJ....122.2267E}.

\section{Data Compression}
\label{sec:datacompression}

In this section we present different types of data compression for analysing galaxy surveys in various scenarios.  In Sec. \ref{subsec:weicorr} we derive optimal weights to be applied to correlation functions, and in Sec. \ref{subsec:weigal} we explore how these weights may be applied to individual galaxies using an ``effective scale" approximation.  These optimal weighting schemes extend to configuration space our previous work \citep{2020MNRAS.498.2948R} for lensing and clustering optimal compression in Fourier space. 

\subsection{Weights applied to correlation functions }
\label{subsec:weicorr}

One of the aims of our work is to derive weight functions able to compress  galaxy clustering and galaxy-galaxy lensing measurements, across different redshifts $z$ and angles $\theta$, with minimal information loss.   To do this, we determine optimal redshift weights for the average tangential shear $\gamma_t(\theta, z)$ and angular galaxy correlation $\omega(\theta, z)$. 

For clarity before proceeding further, we first summarise the optimal weights derived in the case of a single data vector and parameter.  Given an initial dataset $\mathbf{x}$ with dimension $n$, we can linearly compress it into a single number $y$, 
\begin{equation}\label{eq:lincompr}
    y = \mathbf{w}^{\mathrm{T} } \mathbf{x} ,
\end{equation}
which contains the same information as $\mathbf{x}$, with respect to a single parameter $\lambda_i$, if $\mathbf{w} $ is an optimal weight function defined as,
\begin{equation}
\mathbf{w}^{\mathrm{T} }  = \mathbf{C}^{-1 } \mathbf{d} ,
\label{eq:1dweights}
\end{equation}
where $\mathbf{C}$ is the covariance of the data and $\mathbf{d}$ indicates the derivative of the model $\muu$ with respect to the parameter $\lambda_i$ at each point, $\mathbf{d} =  \partial \muu /\partial \lambda_i$. 
 \rossi{In this work we focus on linear compression, which is optimal under the assumption of Gaussianity,  alternative non-linear optimal compression options for Eq. \ref{eq:lincompr} have been discussed in  \cite{2018MNRAS.476L..60A, 2021MNRAS.501..954J}. }
As discussed in \cite{2017MNRAS.464.2698R} the compression is optimal under the assumptions that $\muu$ is Gaussian and that the covariance matrix does not depend on the parameters of interest, i.e. $\partial \mathbf{C}/\partial \lambda_i = 0$.  Although covariance generally does depend on these parameters, it is usually fixed for the purpose of cosmological analyses. Recent lensing-clustering analyses have shown that the cosmology dependence in the covariance matrix affect the parameter constraints at the level of $0.1-0.2 \sigma$
\citep{2021MNRAS.508.3125F, 2021A&A...646A.129J, 2019OJAp....2E...3K}.
\rossi{The covariances assumed here are for Gaussian cases, which means that the weights are in principle sub-optimal in the presence of non-Gaussianity. We investigate the impact of these assumptions in Sec. \ref{sec:results}, where we compare the results of the compressed and uncompressed datasets.}
We now discuss optimal weighting for a multi-dimensional parameter space. We use the example of a combined-probe analysis of galaxy-galaxy lensing and clustering for two parameters $\sigma_8$ and galaxy bias $b$, noting that the same formalism can be applied to different statistics and extended to a larger number of parameters. In particular we derive weights for two scenarios:
\begin{itemize}
    \item[i)] We consider the two statistics measured at a number of redshifts $N_z$ for many scales $N_\theta$, optimally compressed across redshift to the same number of scales, for each of two parameters.  We do not consider the compression of these statistics across scale (i.e.\ to a single scale) because our model is sensitive to scales, i.e. the systematic errors in the model are scale-dependent, and we want to explore this aspect.
    \item[ii)] We study a slightly less-than-optimal data compression involving single-scale weights rather than optimal weights. This is in analogy with FKP weighting \citep{1994ApJ...426...23F} where the weights used are computed fixing a single scale of interest.  In fact, scale-dependent weights cannot be applied to individual galaxies when working in Fourier space or configuration space with standard codes.  We will investigate how sub-optimal  the fixed-scale weights are, compared to scale-dependent optimal weights.
\end{itemize}

For clarity, we set out the full mathematical operations used in these processes as follows \citep[see also][]{2019MNRAS.482.3497Z}:

\vspace{12pt} \noindent {\bf Case i) Two statistics measured at different redshifts for many scales, compressed to many scales for each of 2 parameters.} We construct the  data vector as:
\begin{equation}
\label{eq:uncompdata}
  \text{Data vector } \mathbf{X} = \begin{pmatrix}
    \gamma_t(z_1,\theta_1) \\
    \gamma_t(z_1,\theta_2) \\
    \cdots \\
    \gamma_t(z_2,\theta_1) \\
    \gamma_t(z_2,\theta_2) \\
    \cdots \\
    \omega(z_1,\theta_1) \\
    \omega(z_1,\theta_2) \\
    \cdots \\
    \omega(z_2,\theta_1) \\
    \omega(z_2,\theta_2) \\
    \cdots \\
  \end{pmatrix}
\hspace{1cm} [\text{Dim.} = N_{\rm st}N_zN_\theta \times 1]
\end{equation}
where we also summarise the dimensions of each vector or matrix for clarity. Here, $N_{\rm st}$ denotes the number of statistics used, in our case $N_{\rm st}= 2$.  The covariance matrix $\mathbf{C}$ has dimension $[N_{\rm st} N_zN_\theta \times N_{\rm st} N_zN_\theta]$:
\begin{equation}
\label{eq:uncompcov}
\mathbf{C} =  \begin{pmatrix}
     \mathbf{C}[ {\gamma}_t(\theta_1, z_1) , {\gamma}_t(\theta_1, z_1)]  &   \cdot\cdot &  \mathbf{C}[ {\gamma}_t(\theta_1, z_1) , {\omega}_t(\theta_{m}, z_{n})]    \\
     \cdot\cdot&\cdot\cdot &\cdot\cdot \\
          \mathbf{C}[ \omega(\theta_{m}, z_{n}) , {\gamma}_t(\theta_1, z_1)]  &   \cdot\cdot &  \mathbf{C}[ \omega(\theta_{m}, z_{n}) , {\omega}_t(\theta_{m}, z_{n})] 
  \end{pmatrix}
\end{equation}
with $m = N_{\theta}, n=N_z$, where each term in Eq. \ref{eq:uncompcov} is derived from Eqs. \ref{eq:cov_gamma_T}, \ref{eq:cov_w} or \ref{eq:cross_cov_BA}.  \rossi{ We then construct the  derivative matrix $\textbf{D}$ of dimension $[N_{\rm st} N_zN_\theta \times N_{\rm p} N_\theta]$ as,
\begin{equation}\label{eq:DMulti1}
    D_{[A,n,i],[\alpha,j]} = \frac{\partial \xi_A(\theta_i, z_n)}{\partial p_\alpha} \delta_{ij}
\end{equation}
where $A$ is the statistic, $n$ is the redshift bin, $\alpha$ is the parameter and $(i,j)$ are the separation bins, with $\delta_{ij}$ denoting the Kronecker delta. The full expansion of Eq. \ref{eq:DMulti1} reads,}
\begin{equation}\label{eq:DMULTID}
  \mathbf{D} = \begin{pmatrix}
    \frac{\partial\gamma_t(z_1,\theta_1)}{\partial\sigma_8} & 0 & \cdot & \frac{\partial\gamma_t(z_1,\theta_1)}{\partial b} & 0 & \cdot \\
    0 & \frac{\partial\gamma_t(z_1,\theta_2)}{\partial\sigma_8} & \cdot & 0 & \frac{\partial\gamma_t(z_1,\theta_2)}{\partial b} & \cdot \\
    \cdot & \cdot & \cdot & \cdot & \cdot & \cdot \\
    \frac{\partial\gamma_t(z_2,\theta_1)}{\partial\sigma_8} & 0 & \cdot & \frac{\partial\gamma_t(z_2,\theta_1)}{\partial b} & 0 & \cdot \\
    0 & \frac{\partial\gamma_t(z_2,\theta_2)}{\partial\sigma_8} & \cdot & 0 & \frac{\partial\gamma_t(z_2,\theta_2)}{\partial b} & \cdot \\
    \cdot & \cdot & \cdot & \cdot & \cdot & \cdot \\
    \frac{\partial \omega(z_1,\theta_1)}{\partial\sigma_8} & 0 & \cdot & \frac{\partial \omega(z_1,\theta_1)}{\partial b} & 0 & \cdot \\
    0 & \frac{\partial \omega(z_1,\theta_2)}{\partial\sigma_8} & \cdot & 0 & \frac{\partial \omega(z_1,\theta_2)}{\partial b} & \cdot & \\
    \cdot & \cdot & \cdot & \cdot & \cdot & \cdot \\
    \frac{\partial \omega(z_2,\theta_1)}{\partial\sigma_8} & 0 & \cdot & \frac{\partial \omega(z_2,\theta_1)}{\partial b} & 0 & \cdot \\
    0 & \frac{\partial \omega(z_2,\theta_2)}{\partial\sigma_8} & \cdot & 0 & \frac{\partial \omega(z_2,\theta_2)}{\partial b} & \cdot & \\
    \cdot & \cdot & \cdot & \cdot & \cdot & \cdot \\
  \end{pmatrix}
\end{equation}
where $N_{\rm p}= 2 $ is the number of parameters of interest. We note that $\textbf{D}$ generalizes the 1-parameter derivative in Eq. \ref{eq:1dweights} to a multi-dimensional case. 
\rossi{Each column in Eq. \ref{eq:DMULTID} represents the partial derivative of $\partial \muu_i/ \partial \p$  where $p=\sigma_8,b$, at scale $\theta_i$ with $i = 1 \dots N_\theta$ and redshift $z_n$ with $n = 1 \dots N_z$.
For example, in the case of a single redshift and two different angular bins $N_z = 1, N_\theta =2 $, Eq. \ref{eq:DMULTID} becomes, 
\begin{equation}\label{eq:DMULTIDreduced}
  \mathbf{D} = \begin{pmatrix}
    \frac{\partial\gamma_t(z_1,\theta_1)}{\partial\sigma_8} & 0  & \frac{\partial\gamma_t(z_1,\theta_1)}{\partial b} & 0 & \\
    0 & \frac{\partial\gamma_t(z_1,\theta_2)}{\partial\sigma_8} &  0 & \frac{\partial\gamma_t(z_1,\theta_2)}{\partial b} & \\
    \frac{\partial \omega(z_1,\theta_1)}{\partial\sigma_8} & 0 & \frac{\partial \omega(z_1,\theta_1)}{\partial b} & 0 &  \\
    0 & \frac{\partial \omega(z_1,\theta_2)}{\partial\sigma_8} & 0 & \frac{\partial \omega(z_1,\theta_2)}{\partial b} &  \\
  \end{pmatrix}
\end{equation}
where the $0$ entries are determined by $\delta_{i,j}$ in Eq. \ref{eq:DMulti1}. }
In analogy with Eq. \ref{eq:1dweights}, the optimal weights $\mathbf{W}$ are then constructed as a matrix,
\begin{equation}
\label{eq:weightscale}
 \mathbf{W} = \mathbf{C}^{-1} \, \mathbf{D} \hspace{1cm} [\text{Dim.} = N_{\rm st} N_zN_\theta \times N_{\rm p} N_\theta]
\end{equation}
Applying $\mathbf{W}$ to the initial dataset $ \mathbf{X}$ we obtain the compressed  dataset $\mathbf{Y}$,
\begin{equation}
\label{eq:ycompress}
  \mathbf{Y} = \mathbf{W}^T \, \mathbf{X} \hspace{1cm} [\text{Dim.} = N_{\rm p} N_\theta \times 1]
\end{equation}
and the compressed covariance,
\begin{equation}\label{eq:comprcov}
 \mathbf{C}_Y = \mathbf{W}^T \, \mathbf{C} \, \mathbf{W} \hspace{1cm} [\text{Dim.} = N_{\rm p} N_\theta \times N_{\rm p} N_\theta]
\end{equation}
We note that the compressed statistic $\mathbf{Y}$ contains the same information with respect to the parameters as $\mathbf{X}$, in the sense that the Fisher matrix of $\mathbf{X}$ with respect to the parameters is the same as the Fisher matrix of $\mathbf{Y}$.

\vspace{12pt} \noindent {\bf Case ii): Two statistics measured at different redshifts for many scales, compressed to many scales for each of 2 parameters, using weights evaluated at a single scale.}. This is a slightly less-than-optimal data compression scheme, but by fixing an effective scale it is possible to apply the weights to individual galaxies as we discuss below.  We use the same uncompressed data vector $\mathbf{X}$ and covariance $\mathbf{C}$ as in Case i), but the weights are derived for a single scale $\theta_s$, using the one-dimensional compression (Eq. \ref{eq:1dweights}). We use a single-scale covariance $\mathbf{C}_\mathrm{s}$ and a single-scale derivative matrix $\mathbf{D}_\mathrm{s}$ obtained by evaluating $\mathbf{C}$ and $\mathbf{D}$ at a single scale, \rossi{ e.g. for $\mathbf{D}_\mathrm{s}$ in Eq. \ref{eq:DMulti1} we get
\begin{equation}
%\label{eq:DMulti1}
    D_{[A,n,i],[\alpha,j]} = \frac{\partial \xi_A(\theta_s, z_n)}{\partial p_\alpha} \delta_{ij} 
\end{equation}
}
and 
\begin{equation}\begin{split}
  &\mathbf{D}_\mathrm{s} =\\
  &\begin{pmatrix}
    \frac{\partial\gamma_t(z_1,\theta_s)}{\partial\sigma_8} & 0 & \cdot & \frac{\partial\gamma_t(z_1,\theta_s)}{\partial b} & 0 & \cdot \\
    0 & \frac{\partial\gamma_t(z_1,\theta_s)}{\partial\sigma_8} & \cdot & 0 & \frac{\partial\gamma_t(z_1,\theta_s)}{\partial b} & \cdot \\
    \cdot & \cdot & \cdot & \cdot & \cdot & \cdot \\
    \frac{\partial\gamma_t(z_2,\theta_s)}{\partial\sigma_8} & 0 & \cdot & \frac{\partial\gamma_t(z_2,\theta_s)}{\partial b} & 0 & \cdot \\
    0 & \frac{\partial\gamma_t(z_2,\theta_s)}{\partial\sigma_8} & \cdot & 0 & \frac{\partial\gamma_t(z_2,\theta_s)}{\partial b} & \cdot \\
    \cdot & \cdot & \cdot & \cdot & \cdot & \cdot \\
    \frac{\partial \omega(z_1,\theta_s)}{\partial\sigma_8} & 0 & \cdot & \frac{\partial \omega(z_1,\theta_s)}{\partial b} & 0 & \cdot \\
    0 & \frac{\partial \omega(z_1,\theta_s)}{\partial\sigma_8} & \cdot & 0 & \frac{\partial \omega(z_1,\theta_s)}{\partial b} & \cdot & \\
    \cdot & \cdot & \cdot & \cdot & \cdot & \cdot \\
    \frac{\partial \omega(z_2,\theta_s)}{\partial\sigma_8} & 0 & \cdot & \frac{\partial \omega(z_2,\theta_s)}{\partial b} & 0 & \cdot \\
    0 & \frac{\partial \omega(z_2,\theta_s)}{\partial\sigma_8} & \cdot & 0 & \frac{\partial \omega(z_2,\theta_s)}{\partial b} & \cdot & \\
    \cdot & \cdot & \cdot & \cdot & \cdot & \cdot \\
  \end{pmatrix}
  \end{split}
\end{equation}
and then,
\begin{equation}
  \mathbf{W}_\mathrm{s} = \mathbf{C}_{s}^{-1} \, \mathbf{D}_{s} \hspace{1cm}
         [\text{Dim.} = N_{\rm st} N_z \times N_{\rm p}]
\end{equation}
This weight matrix for a single angular scale $\theta_s$ may be written out more fully as:
\begin{equation}
  \mathbf{W}_{ss} = \begin{pmatrix} W_{\gamma_t,\sigma_8}(z_1,\theta_\mathrm{s}) & W_{\gamma_t,b}(z_1,\theta_\mathrm{s}) \\
    W_{\gamma_t,\sigma_8}(z_2,\theta_\mathrm{s}) & W_{\gamma_t,b}(z_2,\theta_\mathrm{s}) \\
    \cdots & \cdots \\
    W_{w,\sigma_8}(z_1,\theta_\mathrm{s}) & W_{w,b}(z_1,\theta_\mathrm{s}) \\
    W_{w,\sigma_8}(z_2,\theta_\mathrm{s}) & W_{w,b}(z_2,\theta_\mathrm{s}) \\
    \cdots & \cdots \\
  \end{pmatrix}
\end{equation}
\rossi{
These weights are then applied to all scales as an ``approximate''
weight matrix (i.e.\ an approximation to the optimal weights).  This
is achieved by replicating each element of $\mathbf{W}_\mathrm{ss}$ into an
$N_\theta \times N_\theta$ block in order to apply the same weight to
all scales:
\begin{equation}
 %\begin{split}   
 \label{eq:weightapprox}
              W_\mathrm{app} = \begin{pmatrix}
               & W_{\gamma_t,\sigma_8 }(z_1,\theta_\mathrm{s}) \mathcal{I}_\mathrm{i,j}      & W_{\gamma_t,b }(z_1,\theta_\mathrm{s}) \mathcal{I}_\mathrm{i,j} \;\;\\
                &  \cdot    &\cdot                                                                                                      \;  \;\\
                &  \cdot    &\cdot                                                                                                        \;\;\\
                & W_{\gamma_t,\sigma_8}(z_N,\theta_\mathrm{s}) \mathcal{I}_\mathrm{i,j}   & W_{\gamma_t,b}(z_N,\theta_\mathrm{s}) \mathcal{I}_\mathrm{i,j}           \;  \;\\
                & W_{\omega,\sigma_8}(z_1,\theta_\mathrm{s}) \mathcal{I}_\mathrm{i,j}   & W_{\omega,b}(z_1,\theta_\mathrm{s}) \mathcal{I}_\mathrm{i,j}             \\
          %      &W_{\omega,\sigma_8}(z_2,\theta_\mathrm{s}) \mathcal{I}_\mathrm{i,j}   & W_{\omega,b}(z_2,\theta_\mathrm{s}) \mathcal{I}_\mathrm{i,j}             \\
                 & \cdot    &\cdot                                                                                                       \; \;\\
                 &  \cdot    &\cdot                                                                                                       \; \;\\
                 & W_{\omega,\sigma_8}(z_N,\theta_\mathrm{s}) \mathcal{I}_\mathrm{i,j}    & W_{\omega,b}(z_N,\theta_\mathrm{s}) \mathcal{I}_\mathrm{i,j}            \;\\
                \end{pmatrix}
%\end{split}
\end{equation}
where $\mathbf{W}_\mathrm{app}$ has dimension $N_{\rm st}N_z N_\theta \times N_{\rm p}N_\theta$ and $\mathcal{I}_\mathrm{i,j}$ is the identity matrix of dimension $N_\theta \times N_\theta$.
The compressed statistic and covariance matrix are then derived using this approximate weight matrix:
}
\begin{equation}
\label{eq:apprcomprstat}
  \mathbf{Y}_\mathrm{app} = \mathbf{W}_\mathrm{app}^\mathrm{T} \, \mathbf{X} \hspace{1cm} [\text{Dim.} = N_{\rm p} N_\theta \times 1]
\end{equation}
and,
\begin{equation}
\label{eq:apprcomprcov}
\mathbf{C}_{\mathbf{Y}\mathrm{app}} = \mathbf{W}_\mathrm{app}^\mathrm{T} \,
  \mathbf{C} \, \mathbf{W}_\mathrm{app} \hspace{1cm} [\text{Dim.} =
    N_{\rm p} N_\theta \times N_{\rm p} N_\theta]
\end{equation}
In this case, unlike in Case i), information is lost in the sense that the Fisher matrix of $\mathbf{Y}_\mathrm{app}$ is not the same as the Fisher matrix of $\mathbf{X}$.  We'll compare these different compression schemes as applied to simulated data in Sec. \ref{sec:results} below.

\subsection{Weights applied to individual galaxies}
\label{subsec:weigal}

We now outline a scheme whereby the optimal correlation function weights derived above can be applied directly to individual galaxies in data analysis.  In this scenario the weights can be used for more general applications, outside of data compression. Similarly to the FKP weights, we can use them as optimal weights for lenses when combining galaxy clustering and lensing.

In Sec. \ref{subsec:weicorr} we determined optimal redshift weights for a joint analysis of the angular galaxy correlation function $\omega(\theta)$ and the average tangential shear $\gamma_t(\theta)$ around those galaxies, for creating the compressed statistic $Y(\theta)$.  For each model parameter, we can use Eq. \ref{eq:apprcomprstat} to re-write those relations as:
\begin{eqnarray}
\label{eq:corrpergalaxy}
  \omega_{\mathrm{tot}}(\theta) &=& \sum_i W_{\omega,i} \, \omega_i(\theta) , \\
  \gamma_{t, \mathrm{tot}}(\theta) &=& \sum_i W_{\gamma,i} \, \gamma_{t,i}(\theta) , \\
  Y(\theta) &=& \omega_{\mathrm{tot}}(\theta) + \gamma_{t, \mathrm{tot}}(\theta) ,
\end{eqnarray}
where $i$ labels the redshift bins, $W_{\omega,i}$ is the optimal redshift weight for $\omega(\theta)$ in redshift bin $i$, and $W_{\gamma,i}$ is the optimal redshift weight for $\gamma_t(\theta)$, where those weights are determined using Eq. \ref{eq:weightapprox}.

In what follows we replicate these combinations by applying weights to each of the individual objects (galaxies and random points) as a function of redshift, and measuring a correlation function across all redshifts.  We denote  with $g_{\omega,i}$  the per-galaxy weights for $\omega(\theta)$, and with $g_{\gamma,i}$ the per-galaxy weights for $\gamma_t(\theta)$.

For $\omega(\theta)$ we consider an estimator for the total correlation
function across redshifts of the form:
\begin{equation}
\label{eq:weicorrest}
  \omega_{\rm tot}(\theta) = \frac{\widetilde{DD}_{\rm tot} - 2 \,
    \widetilde{DR}_{\rm tot} + \widetilde{RR}_{\rm tot}}{RR_{\rm tot}}
\end{equation}
\citep{2016MNRAS.461.2867Z}, where the numerator involves weighted total pair counts, which (for the purposes of the derivation) we can express in terms of a sum over the pair counts in redshift bins,
\begin{eqnarray}
  \widetilde{DD}_{\rm tot}(\theta) &=& \sum_i \sum_j g_{\omega,i} \,
  g_{\omega,j} \, D_iD_j(\theta) , \\ \widetilde{DR}_{\rm tot}(\theta) &=&
  \sum_i \sum_j g_{\omega,i} \, g_{\omega,j} \, D_iR_j(\theta) ,
  \\ \widetilde{RR}_{\rm tot}(\theta) &=& \sum_i \sum_j g_{\omega,i} \,
  g_{\omega,j} \, R_iR_j(\theta) ,
\end{eqnarray}
where $D_iD_j$, $D_iR_j$, $R_iR_j$ are the auto/cross-pair counts between redshift bins $i$ and $j$, and the denominator of Eq. \ref{eq:weicorrest} involves unweighted total random pair counts.  Hence,
\begin{equation}
  \omega_{\rm tot}(\theta) = \frac{\sum_i \sum_j g_{\omega,i} \, g_{\omega,j} \left( D_iD_j - 2 \, D_iR_j + R_iR_j \right)}{RR_{\rm tot}} .
\end{equation}
Substituting in the auto/cross-correlation function between redshift bins $i$ and $j$, $\omega_{ij} = (D_iD_j - 2 D_iR_j + R_iR_j)/R_iR_j$ we find,
\begin{equation}
  \omega_{\rm tot}(\theta) = \frac{\sum_i \sum_j g_{\omega,i} \, g_{\omega,j} \,
    R_iR_j(\theta) \, \omega_{ij}(\theta)}{RR_{\rm tot}(\theta)}.
\end{equation}
In the approximation that the cross-correlations between different
redshift slices are zero, $\omega_{ij} = 0$ if $i \ne j$:
\begin{equation}
  \omega_{\rm tot}(\theta) = \frac{\sum_i g_{\omega,i}^2 \, RR_i(\theta) \,
    \omega_i(\theta)}{RR_{\rm tot}(\theta)} ,
\end{equation}
where we use $RR_i = R_iR_i$.  Comparing the form of this equation with Eq. \ref{eq:corrpergalaxy}, the galaxy and random weights $g_{\omega,i}$ can be written in terms of the $\omega(\theta)$ weights $W_{\omega,i}$ in redshift bins $i$:
\begin{equation}
\label{eq:wpergalaxy}
  g_{\omega,i} = \sqrt{ \frac{W_{\omega,i} \, RR_{\rm tot}(\theta_{\rm
        eff})}{RR_i(\theta_{\rm eff})} } ,
\end{equation}
evaluated at some effective scale $\theta = \theta_{\rm eff}$.

We can derive similar per-galaxy weights for $\gamma_\mathrm{t}(\theta)$. For the purposes of this derivation we'll assume an estimator of the form:
\begin{equation}
  \gamma_t(\theta) = \frac{\sum_{ls} w_l \, w_s \, e_{t,ls}}{\sum_{ls}
    w_l \, w_s} ,
\end{equation}
where $w_l$ and $w_s$ are lens weights and source weights, and $e^t_{ls}$ indicates the tangential shear of sources around lenses.  (This estimator excludes the random lens term, but this term makes no contribution to the signal.)  We further assume $w_s = 1$ for the purposes of this derivation. \rossi{Similarly to Eq. \ref{eq:weicorrest}, we apply $w_l = g_{\gamma_\mathrm{t},i}$ in the numerator of the estimator and $w_l = 1$ in the denominator, such that},
\begin{equation}
  \gamma_{\mathrm{tot}}(\theta) = \frac{\sum_{ls} g_{\gamma,i} \,
    e_{t,ls}}{D_{\rm tot}D_s} ,
\end{equation}
where $D_{\rm tot}D_s$ is the cross-pair count between all the lenses and the sources.  Breaking the numerator into tangential shear measurements in different redshift bins, $\gamma_i = \sum_{ls} e_{t,ls}/D_i D_s$, we find:
\begin{equation}
  \gamma_{\rm tot}(\theta) = \frac{\sum_i g_{\gamma,i} \,
    D_iD_s(\theta) \, \gamma_i(\theta)}{D_{\rm tot}D_s(\theta)} .
\end{equation}
Comparing the form of this equation with Eq. \ref{eq:corrpergalaxy}, we can find the lens galaxy weights $g_{\gamma,i}$ in terms of the $\gamma_t(\theta)$ weights $W_{\gamma,i}$ in redshift bins $i$:
\begin{equation}
\label{eq:gtpergalaxy}
  g_{\gamma,i} = \frac{W_{\gamma,i} \, D_{\rm tot}D_s(\theta_{\rm eff})}{D_iD_s(\theta_{\rm eff})} .
\end{equation}
Eqs. \ref{eq:wpergalaxy} and \ref{eq:gtpergalaxy} hence give the individual galaxy redshift weights for measurements of $\omega(\theta)$ and $\gamma_t(\theta)$, which are equivalent to the optimal correlation function weights derived in Case ii) of Sec. \ref{subsec:weicorr}.

\section{Simulated data and measurements}
\label{sec:measurements}

\subsection{Simulations}
\label{sec:simulation}

We tested these algorithms using simulated catalogues that were representative of future overlapping DESI and weak lensing datasets.  We sampled these catalogues from the Buzzard simulation suite \citep{2019arXiv190102401D}, which uses full N-body particle simulations to derive lightcones, dark matter halo catalogues and lensing fields via ray-tracing.  The Buzzard lightcones are populated with a complete sample of galaxies to apparent magnitude limit $i = 27$, using the ADDGALS \citep{2021arXiv210512105W} abundance-matching algorithm to sample luminosities and colours.  Convergence and shear fields are derived using the CALCLENS \citep{2013MNRAS.435..115B} algorithm, and shear components are assigned to each simulated galaxy.

We selected lens and source samples from these Buzzard lightcone catalogues. First, using the simulated galaxy magnitudes, we created DESI Bright Galaxy (BGs) and Luminous Red Galaxy (LRGs) targets matching the densities and redshift distributions of these samples.  We produced a total lens sample across the redshift range $0.1 < z < 0.9$ by combining BGs in the range $0.1 < z < 0.5$, with LRGs in the range $0.5 < z < 0.9$.  We then created a representative weak lensing source sample of angular density 10 arcmin$^{-2}$ by sub-sampling the Buzzard shear catalogues with a source redshift distribution
\begin{equation}
  p(z) \propto z^2 e^{-z/z_0} ,
\end{equation}
where $z_0 = 0.2$.  We assigned a shape noise error of $\sigma_e = 0.28$ per ellipticity component, and a photometric redshift error of $\sigma_z = 0.1 \, (1 + z)$.  For the purposes of our current study, we did not simulate any multiplicative shape bias corrections and we assumed unity weights for completeness and shape measurement for all sources and lenses.

Each Buzzard lightcone is built across half the sky.  We used one of these half-sky catalogues, and divided it into 18 pseudo-independent regions each of area $\sim 1000$ sq deg, which is representative of the overlap area between DESI and each of the weak lensing datasets DES, HSC and KiDS.  We analysed each of these 18 regions separately, to study the sample variance in our results.

\subsection{Correlation function measurements}

We measured the correlation functions of the simulated catalogues for each redshift bin and angular scale.  For the angular correlation function we used the Landy-Szalay estimator \citep{1993ApJ...412...64L},
\begin{equation}
\label{eq:wtestimator}
\omega = \frac{DD - 2DR + RR }{RR} ,
\end{equation}
for each angular scale $\theta$. For $\gamma_t$ measurements we used the estimator,
\begin{equation}\label{eq:gamtestimator}
    \gamma_t = \frac{(N_R/N_D)\sum_{ls} e_{t,ls} - \sum_{rs} e_{t,rs} }{R_r R_s} ,
\end{equation}
where $N_D$ and $N_R$ are the number of data and random points, and $R_rR_s$ denotes the cross-pair count between the random lenses in each redshift bin and all sources.  We computed the correlation functions described in Eqs. \ref{eq:wtestimator} and \ref{eq:gamtestimator} using the public code \textit{treecorr} \citep{2015ascl.soft08007J}.  For this analysis we used 30 logarithmically-spaced angular separation bins in the range $0.003<\theta< 3$ deg, and 8 linearly-spaced lens redshift bins in the range $0.1 < z < 0.9$.

\begin{figure*}
\begin{minipage}{\textwidth}
\centering
 % \begin{multicols}[3]
    \includegraphics[width= 0.3\columnwidth]{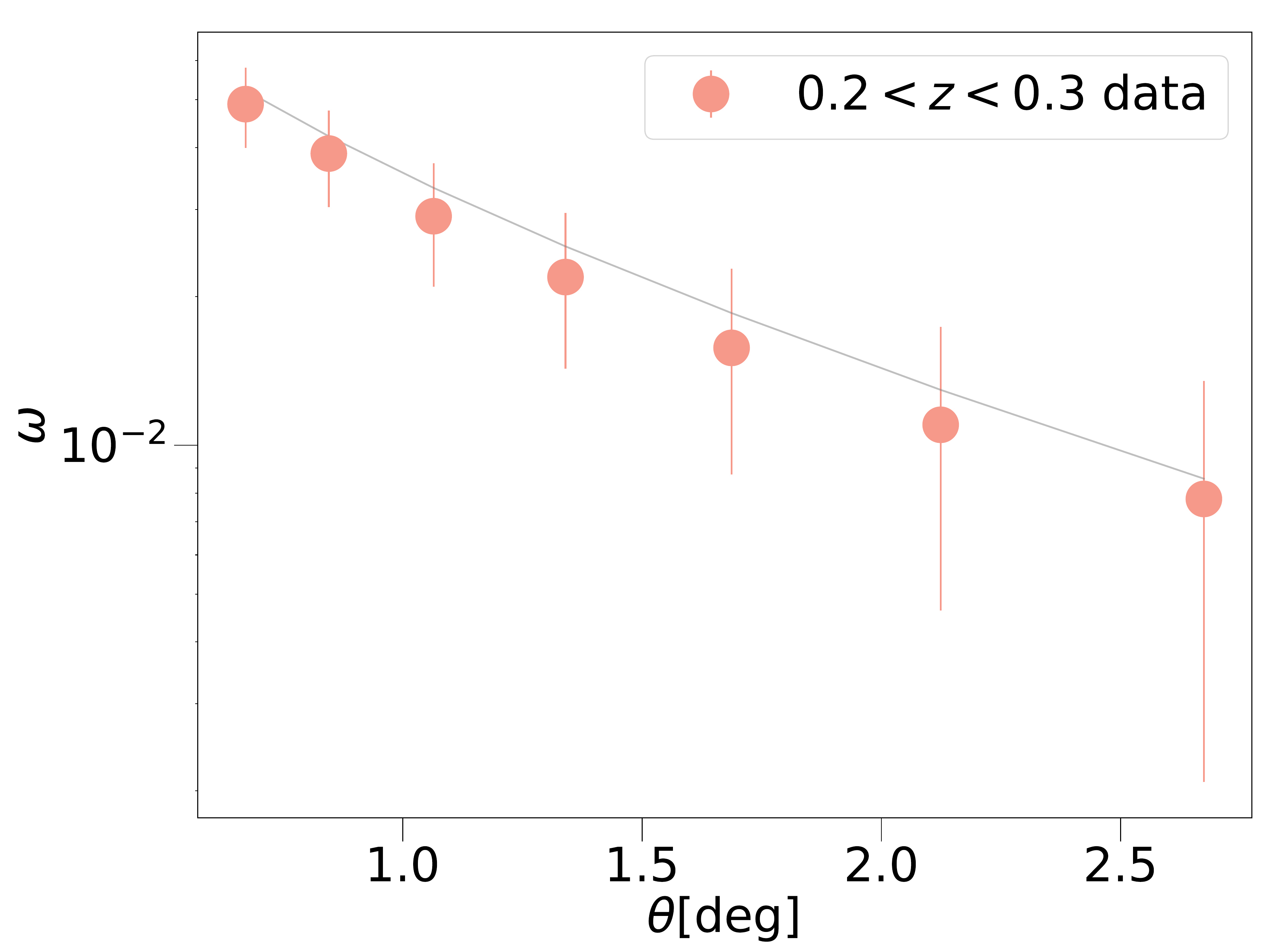}
    \includegraphics[width= 0.3\columnwidth]{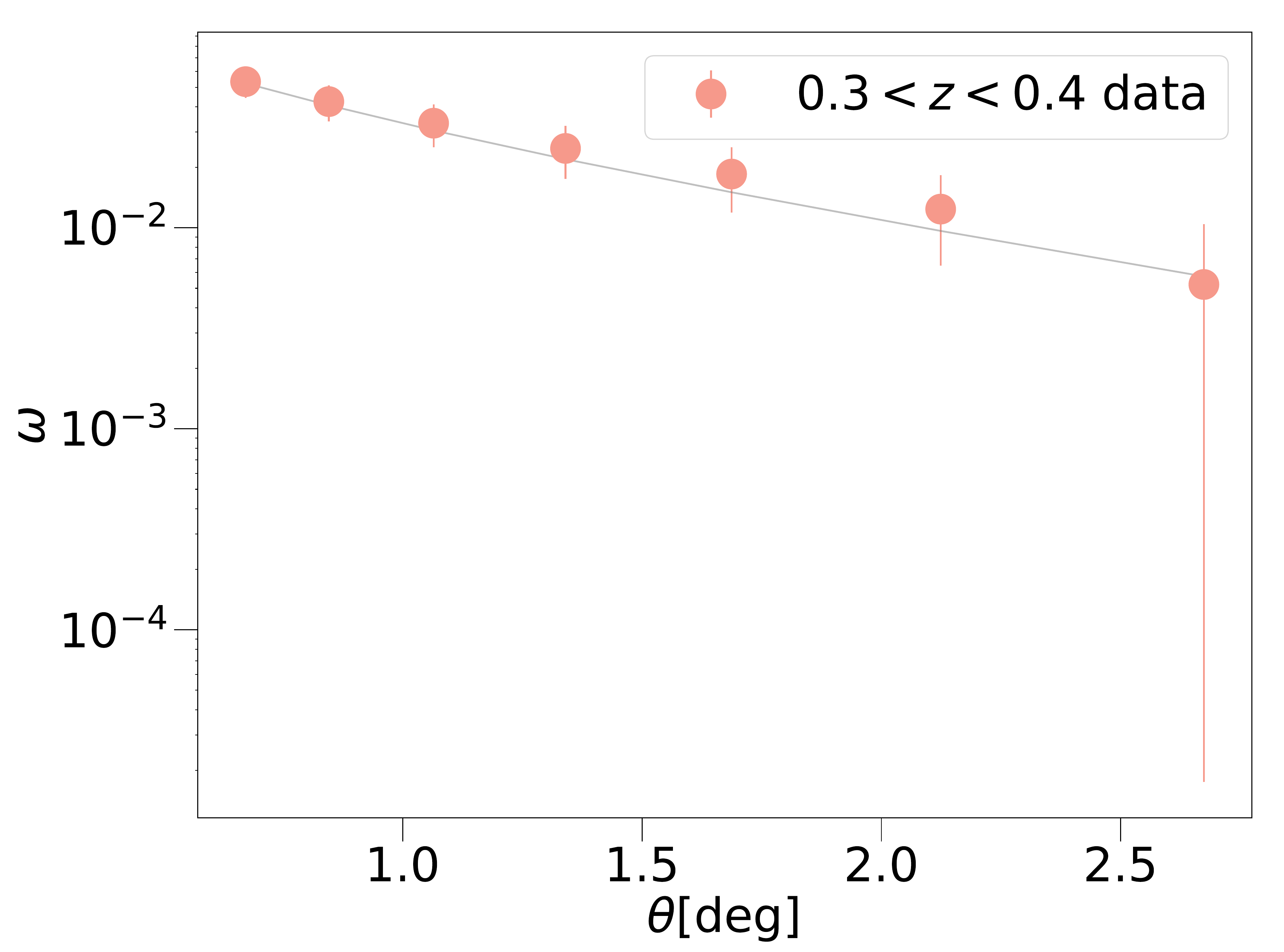}
    \includegraphics[width= 0.3\columnwidth]{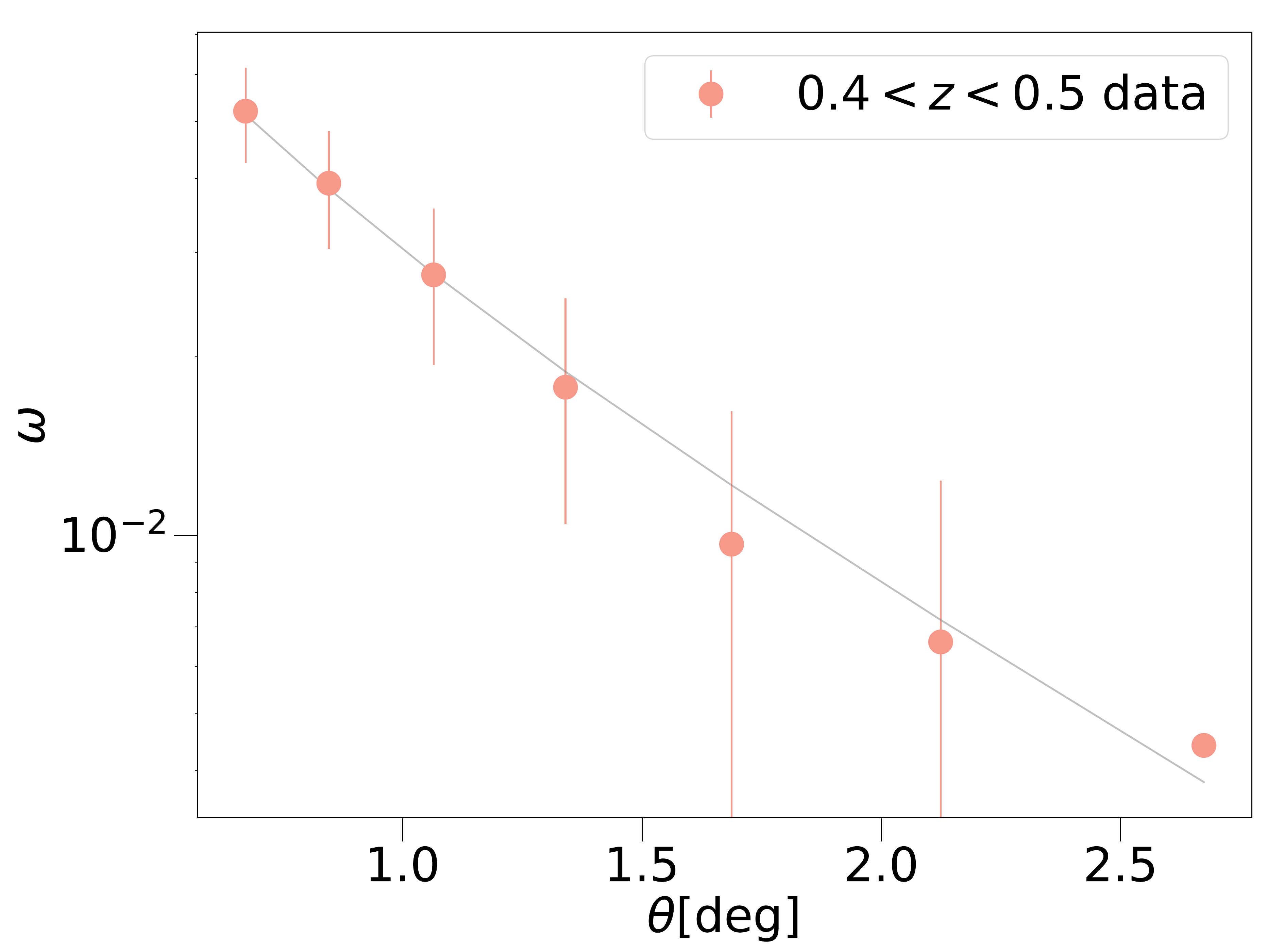}
    \includegraphics[width= 0.32\columnwidth]{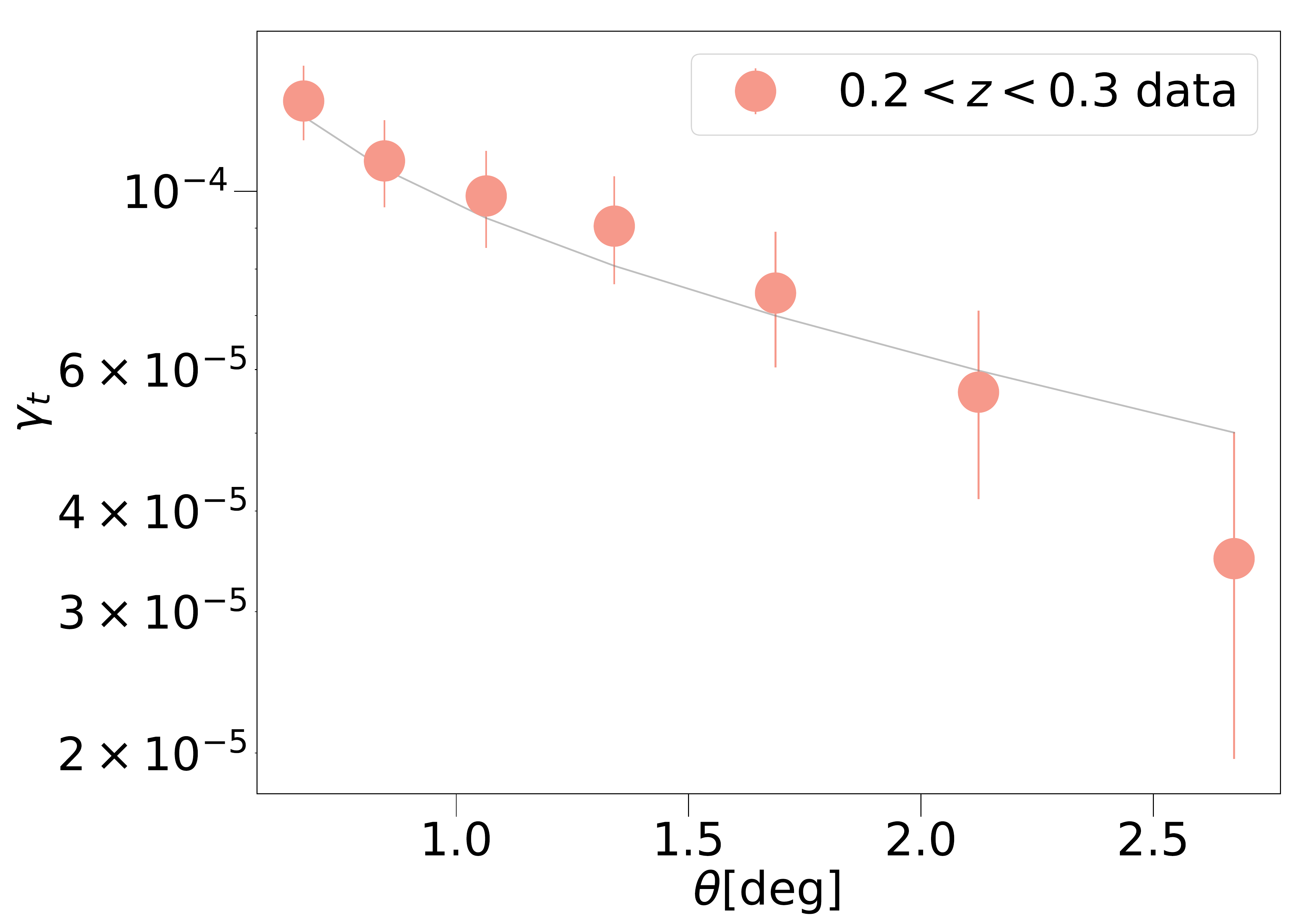}
    \includegraphics[width= 0.32\columnwidth]{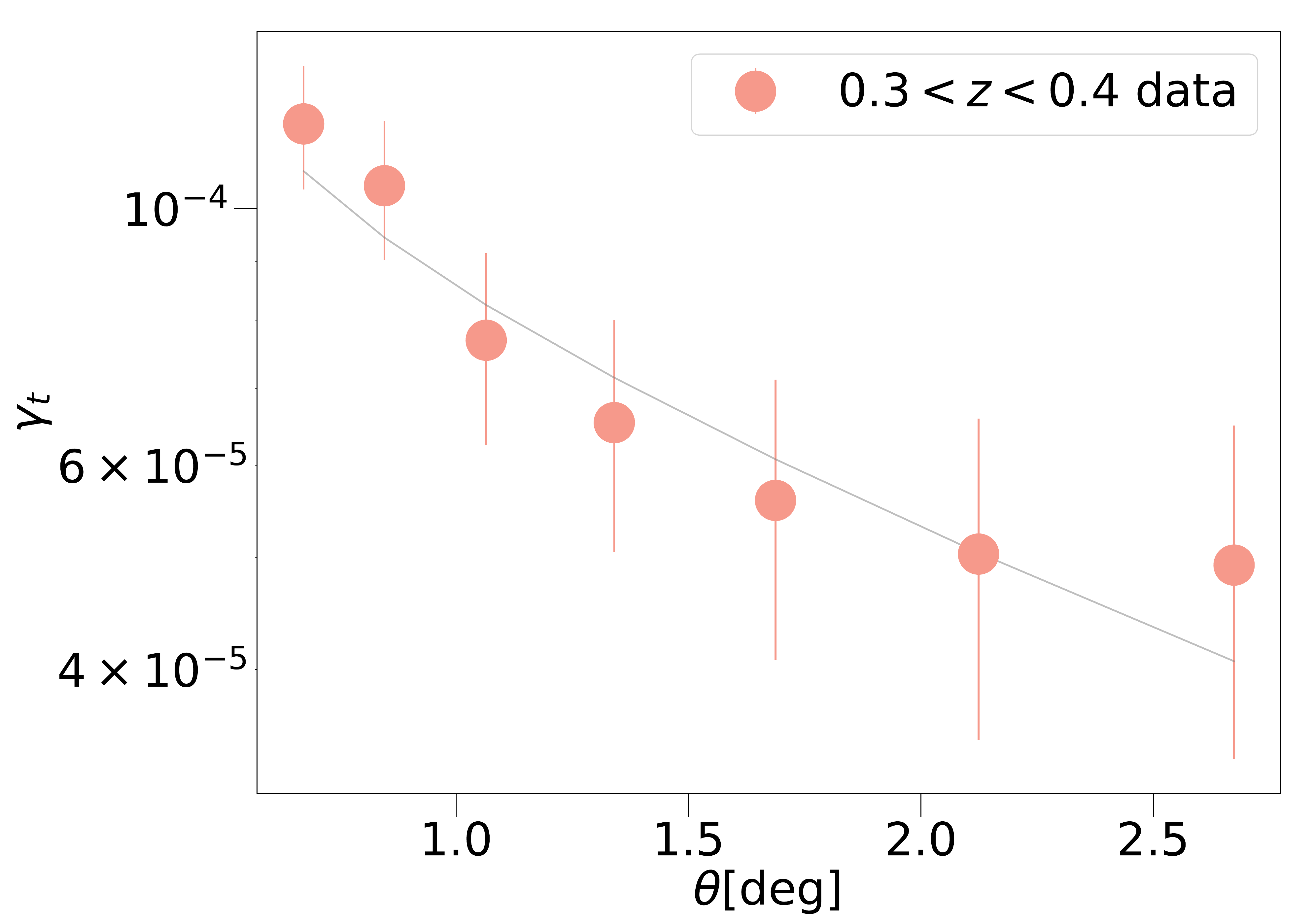}
    \includegraphics[width= 0.32\columnwidth]{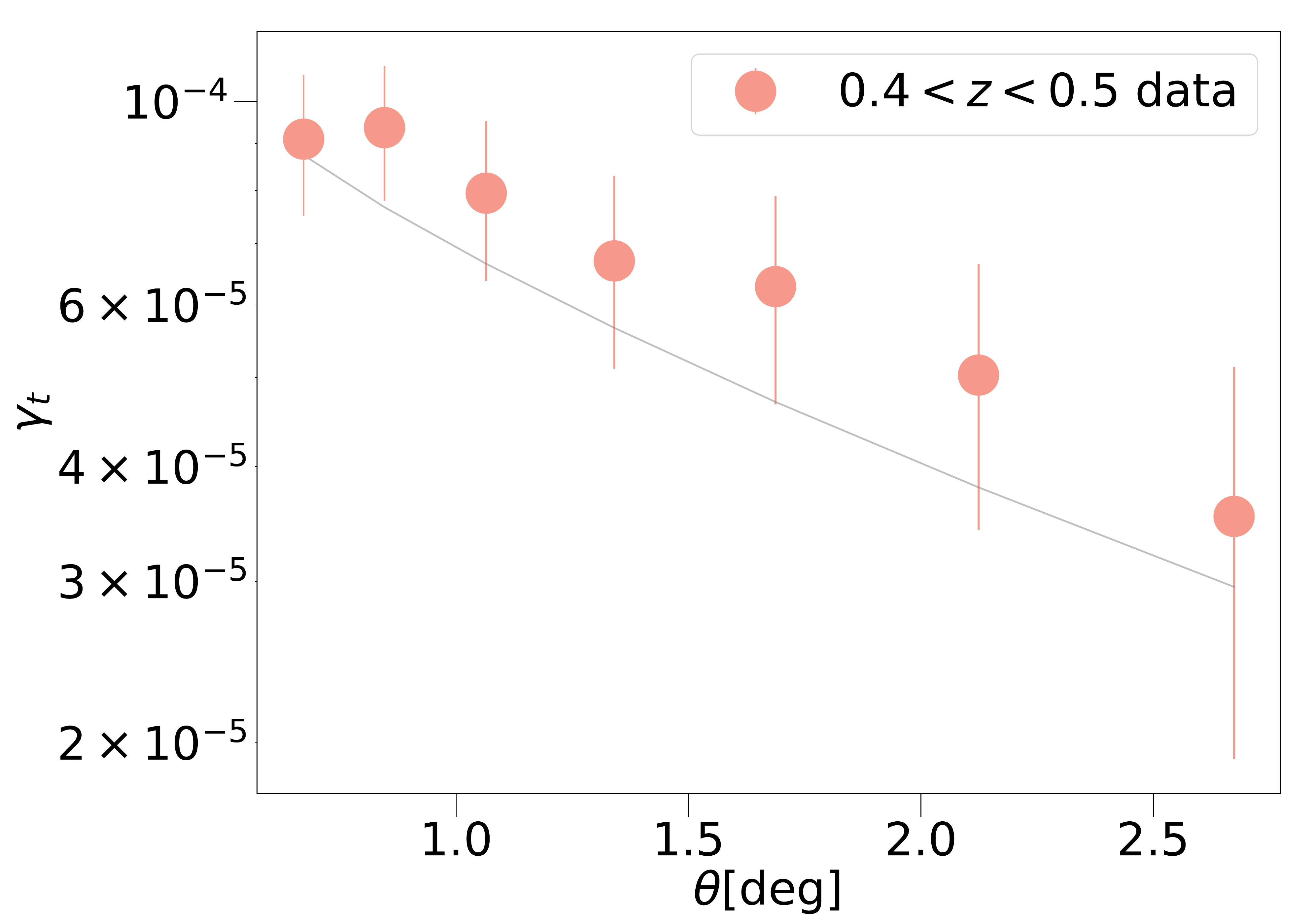}
 %\end{multicols}
    \end{minipage}
  \caption{
    The angular galaxy correlation function $\omega(\theta)$ (top panels) and the average tangential shear $\gamma_t(\theta)$ (bottom panels), \rossi{measured on one region, }across three of the eight redshift bins between $0.2 < z < 0.3$, $0.3 < z < 0.4$, $0.4 < z < 0.5$  with scale cut $R(\theta) > 5 h^{-1}$ Mpc. For display purposes, we plot the minimum common fitted $\theta$ range across for all redshifts.  The best fit model is plotted in grey for all redshift bins. Error bars are computed from the diagonal of the covariance matrix.}
   \label{fig:wt_gt_zbin}
\end{figure*}

\begin{figure}
    \centering
    \includegraphics[width= \columnwidth]{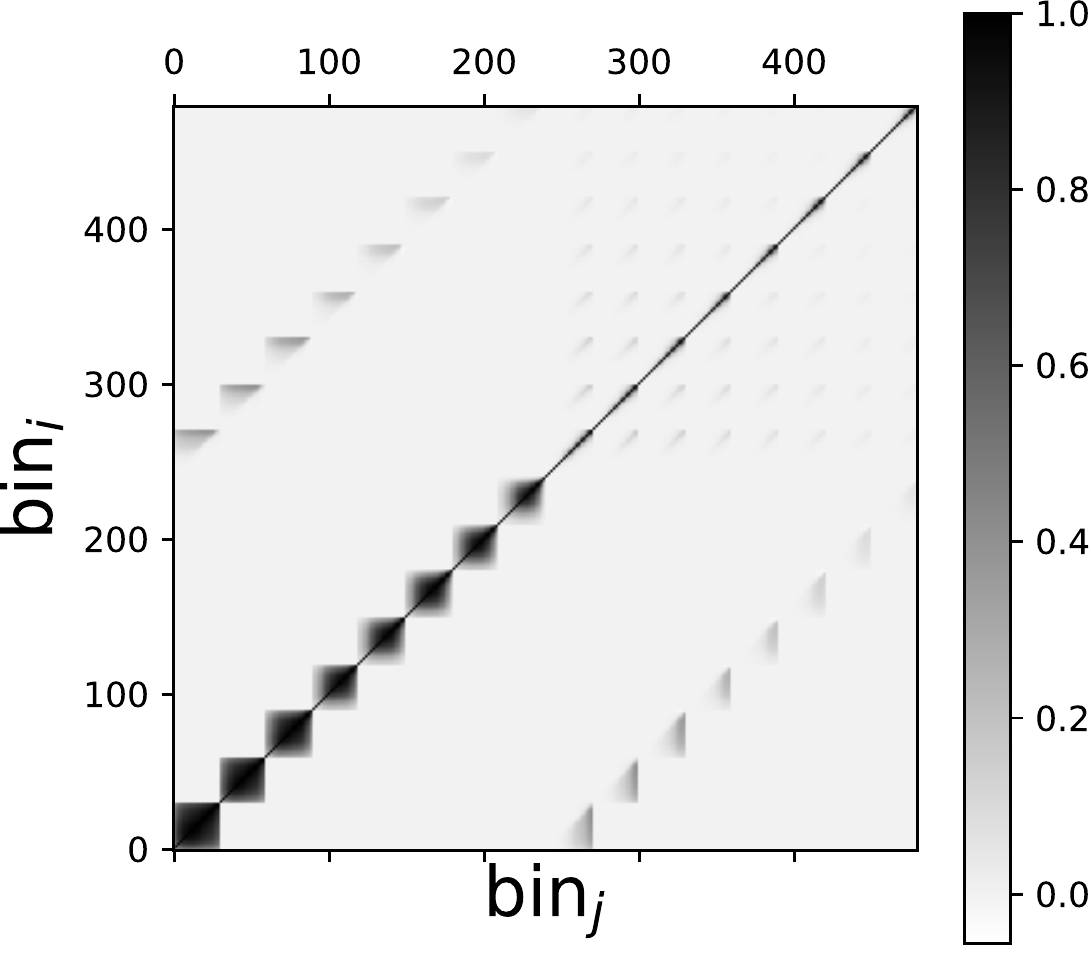}
    \caption{The correlation matrix of the uncompressed dataset. The bins are ordered by statistic, redshift and scale, with $i$ and $j$ indices ranging from 1 to $N_\mathrm{st} \times N_z \times N_\theta$ as illustrated in Eq. \ref{eq:uncompcov}, with $N_\mathrm{st}=2 $ for this case.}
    \label{fig:covariance_uncompressed}
\end{figure}

Figure \ref{fig:wt_gt_zbin} displays the uncompressed correlation functions across three of the eight redshift bins (red circles)  $0.2 < z < 0.3$, $0.3 < z < 0.4$, $0.4 < z < 0.5$, measured from one of the Buzzard regions.  The top panel corresponds to the measurements of $\omega(\theta)$, while the bottom panel indicates the measurements of $\gamma_t(\theta)$. The best-fit model is plotted as grey continuous lines.
For display purposes we plot the minimum common fitted $\theta$ range across for all redshifts (see Sec.\ \ref{subsec:redweights}). 
The error-bars are derived as the square root of the diagonal covariance matrix elements defined in Sec. \ref{sec:cov}.  The full covariance of the uncompressed statistics is displayed as a correlation matrix in Fig. \ref{fig:covariance_uncompressed}.

\section{Results}
\label{sec:results}

We used the lensing and galaxy clustering statistics defined in Sec. \ref{sec:model} and measured in Sec. \ref{sec:simulation}, and fit for the $\sigma_8$ and galaxy bias parameters using several different approaches.  In Sec. \ref{subsec:uncompfit} we present the results obtained using a standard approach fitting to the uncompressed measurements, where a tomographic analysis is performed across the full redshift range.  In Sec. \ref{subsec:redweights} we review the optimal redshift weights obtained in this analysis, which may be used to compute individual galaxy weights, and in Sec. \ref{subsec:compfit} we present the results obtained from data-compressed analyses, using the methodology described in Sec. \ref{sec:datacompression}.  We compare and contrast the outcomes of these different approaches in Sec. \ref{subsec:compare}.

We fit 18 individual Buzzard simulation regions, and present the average and dispersion of the results.  All fits are performed at angular scales where $R(\theta) > 5 \, h^{-1} \mathrm{Mpc}$, where $R$ is the projected co-moving galaxy separation at the redshift of the measurement, such that a linear galaxy bias model applied to an acceptable approximation given our context.  We repeated the analysis using a different scale cut $R(\theta) > 8 \, h^{-1} \mathrm{Mpc}$, finding consistent results (see Appendix \ref{sec:appendix}.

We sampled the posterior distributions, using BILBY \citep{2019ApJS..241...27A}. We ran 8 independent chains for each fit, each consisting of 500 walkers. We determined the convergence of the samples using the Gelman-Rubin criteria (R-1 <0.01) \citep{Gelman:1992zz}. The flat priors used on the parameters are specified in each individual analysis below.
For simplicity of notation we present the different analyses under the labels of Case i)-ii)-iii)-iv) which indicate:
\begin{itemize}
    \item[ i)]  Data compression of $\gamma_t$ and $\omega$ using scale and redshift dependent weights and their covariance. We compress  $\gamma_t$ and $\omega$ with respect to four parameters: $\sigma_8,b_0, b_1, b_2$. 
    The covariance between the statistics is compressed using the same scale-dependent weighting scheme. 
    \item[ ii)]  Data compression of $\gamma_t$ and $\omega$ using redshift dependent weights fixing one single effective scale in the weights. We compress  $\gamma_t$ and $\omega$ with respect to four parameters: $\sigma_8,b_0, b_1, b_2$. 
    The covariance between the statistics is compressed using the same effective-scale  weighting scheme. 
    \item[ iii)]  The uncompressed analysis, considering $\gamma_t$ and $\omega$ across different redshift bins and scales and their covariance. 
    \item[ iv)]  Same as ii) but using a linear model for the bias-redshift relation to investigate possible systematics, i.e. we set $b_2 = 0$ and compress  $\gamma_t$ and $\omega$ with respect to: $\sigma_8,b_0, b_1$. 
\end{itemize}

\subsection{Fits to uncompressed correlations}
\label{subsec:uncompfit}

For the uncompressed analysis, that we refer to as Case iii), we simultaneously fit models to the $N_z = 8$ redshift bins.  We built a data vector $ \mathbf{X}$ combining measurements of $\gamma_t (z, \theta)$ and $\omega(z, \theta)$, following Eq. \ref{eq:uncompdata}, and used the full joint covariance derived from Eq. \ref{eq:uncompcov}.  After applying the scale cut $R(\theta) > 5 \, h^{-1} \mathrm{Mpc}$, we used a remaining number of $87$ data points across all redshifts.  We fit for a total of 9 model parameters: a single value of $\sigma_8$, and one linear bias parameter for each redshift bin: $b_i$ with $i= 1, \dots, 8$.  We set a uniform prior on all the parameters $ \sigma_8, b_i \in [0, 4]$. 

\rossi{Assuming a Gaussian likelihood, the $\chi^2$ statistic used to determine the likelihood of each model is computed as,}
\begin{equation}
    \chi^2 =  \mathbf{(M-X)^\mathrm{T} C^{-1} (M- X) } ,
\end{equation}
where \textbf{M} is the model vector for $\gamma_t (z, \theta)$ and $\omega(z, \theta)$, ordered in the same sequence as $\mathbf{X}$. The $\mathbf{M}$ terms are computed using Eq. \ref{eq:gammatamodel} and \ref{eq:wtetamodel}, and subject to the same scale cuts.
\rossiiii{Note that we re-generate the covariance with the bias factors matching the best fit for each redshift slice.}

We find $\sigma_{8,\mathrm{mean}} = 0.76 \pm 0.044$, where we quote the mean of the maximum likelihood estimate over 18 regions, and the error as the mean standard deviation of the posterior distribution.  For the minimum value of $\chi^2$, we find a mean value of $\chi_{\mathrm{mean}}^2/\mathrm{d.o.f} = 0.84$, indicating an acceptable goodness-of-fit.  We find a $\sigma_8$ value consistent with, but slightly lower than, the fiducial value of the mocks.  Possible reasons for this include the adoption of a linear galaxy bias model, which may not be fully accurate.  Since the focus of this analysis is on evaluating the error (the information loss) in the parameters, rather than necessarily recovering a particular fiducial value, this issue does not impact our conclusions. We also check that the different fitting cases lead to very similar results in terms of best fits, to ensure that the compression does not add biases.

\begin{figure}
    \centering
\includegraphics[width= \columnwidth]{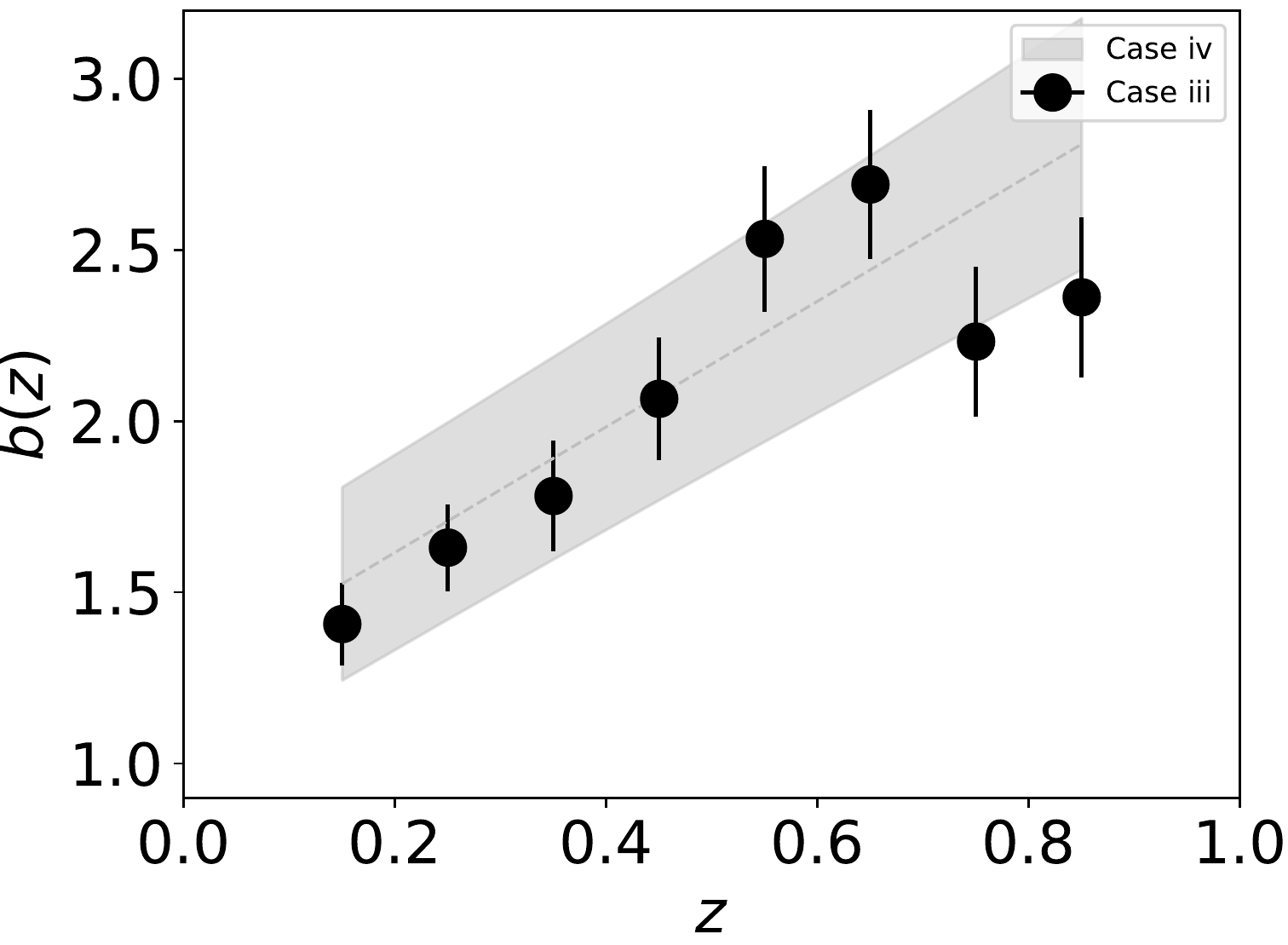}
\includegraphics[width= \columnwidth]{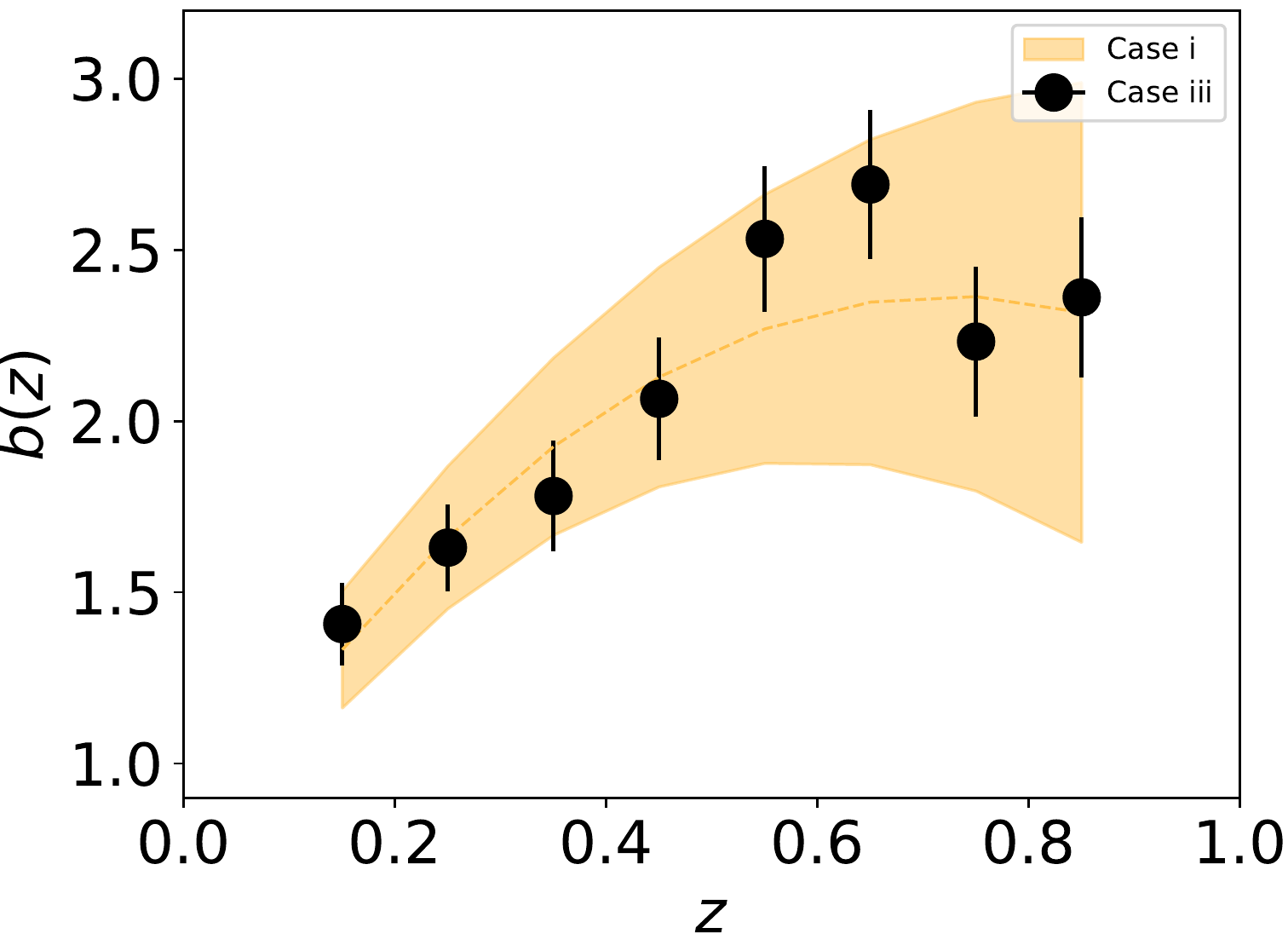}
\includegraphics[width= \columnwidth]{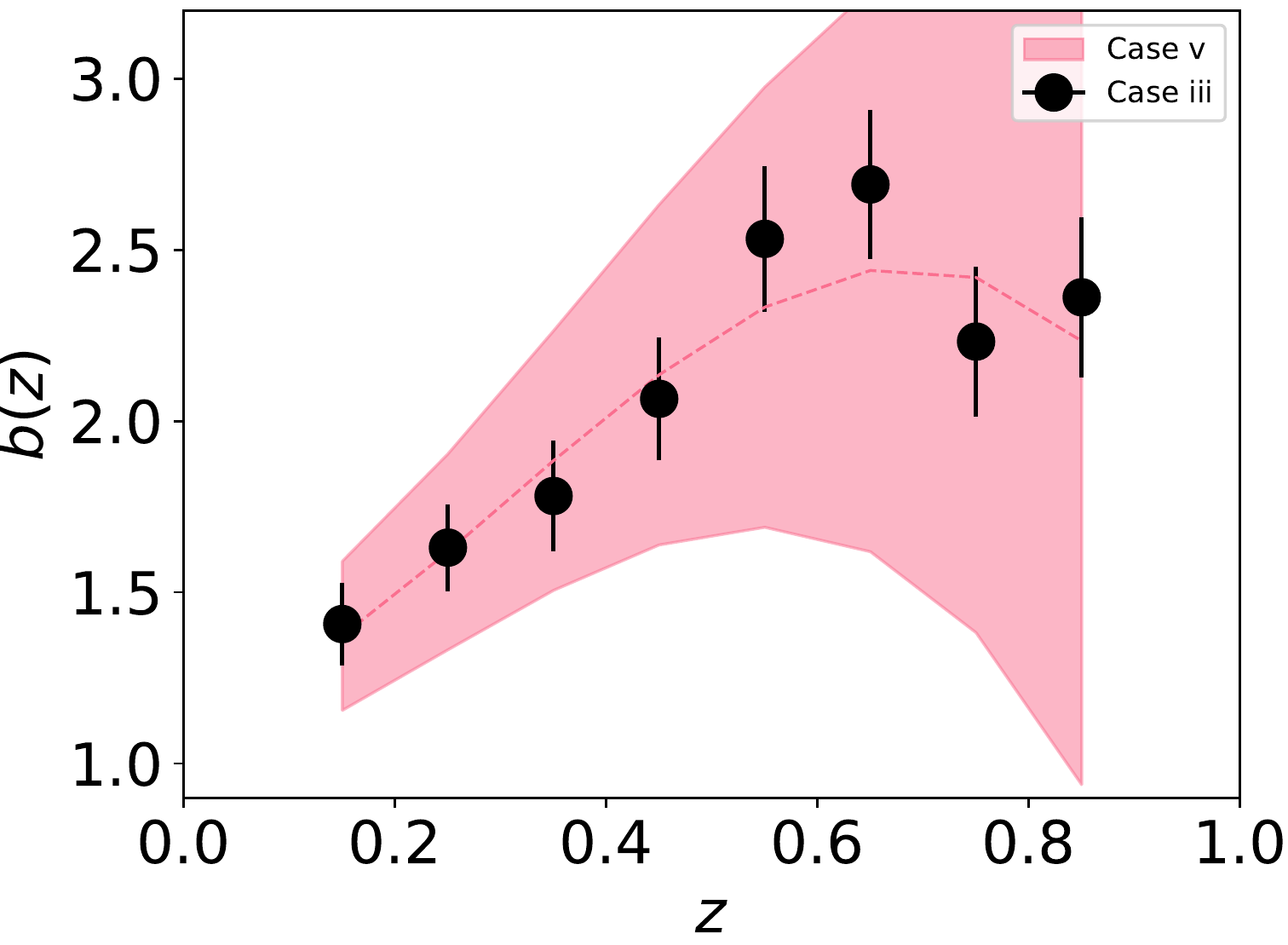}
\caption{Parameter fits for the bias-redshift relation \rossi{for one of the regions analysed},  from the uncompressed analysis (Case iii, black dots in all three panels) and the following compressed analyses:   linear bias-redshift relation where $b_2 = 0$ in Eq.\ref{eq:bias} (Case iv, grey regions, top panel) and quadratic bias (Case i, yellow regions, middle panel) and cubic bias (Case v, red regions, bottom panel) . The errors displayed are computed from the standard deviation of the posterior distribution.}
    \label{fig:bias_quad}
\end{figure}

The results for the bias parameters \rossi{for one of the regions analysed},  for each redshift slice are shown in Fig. \ref{fig:bias_quad} as the turquoise dots, where the error bars indicate the $1\sigma$ ranges.  We find that the galaxy bias increases steadily with redshift as predicted by a hierarchical model of galaxy formation \citep{1996MNRAS.282..347M} and described by \cite{2016arXiv161100036D}. 
 The $b$ versus $z$ trend is dominated by changes in luminosity, as predicted by the increased luminosity of galaxies vs redshift.  As noted in Sec. \ref{subsec:fidcos} we confirm the smooth relationship of $b$ vs $z$ around $z=0.5$, between BG and LRGs samples. 

\subsection{Optimal redshift weights}
\label{subsec:redweights}

We considered two types of data compression, Case i) and Case ii), as described in Sec. \ref{sec:datacompression}, which correspond to optimal correlation function weights and those determined at a single scale, respectively.  The initial dataset \textbf{X} and uncompressed model are computed as for the uncompressed data, and we apply the same scale cut $R(\theta) > 5 \, h^{-1} \mathrm{Mpc}$ as for the uncompressed analysis.  We selected 4 parameters of interest $\sigma_8$ and $b_0, b_1, b_2$, describing a linear galaxy bias with a quadratic evolution in redshift, as described in Sec. \ref{sec:model}, and compared our results with linear and cubic bias evolution choices.

\begin{figure}
    \centering
     \includegraphics[width= \columnwidth]{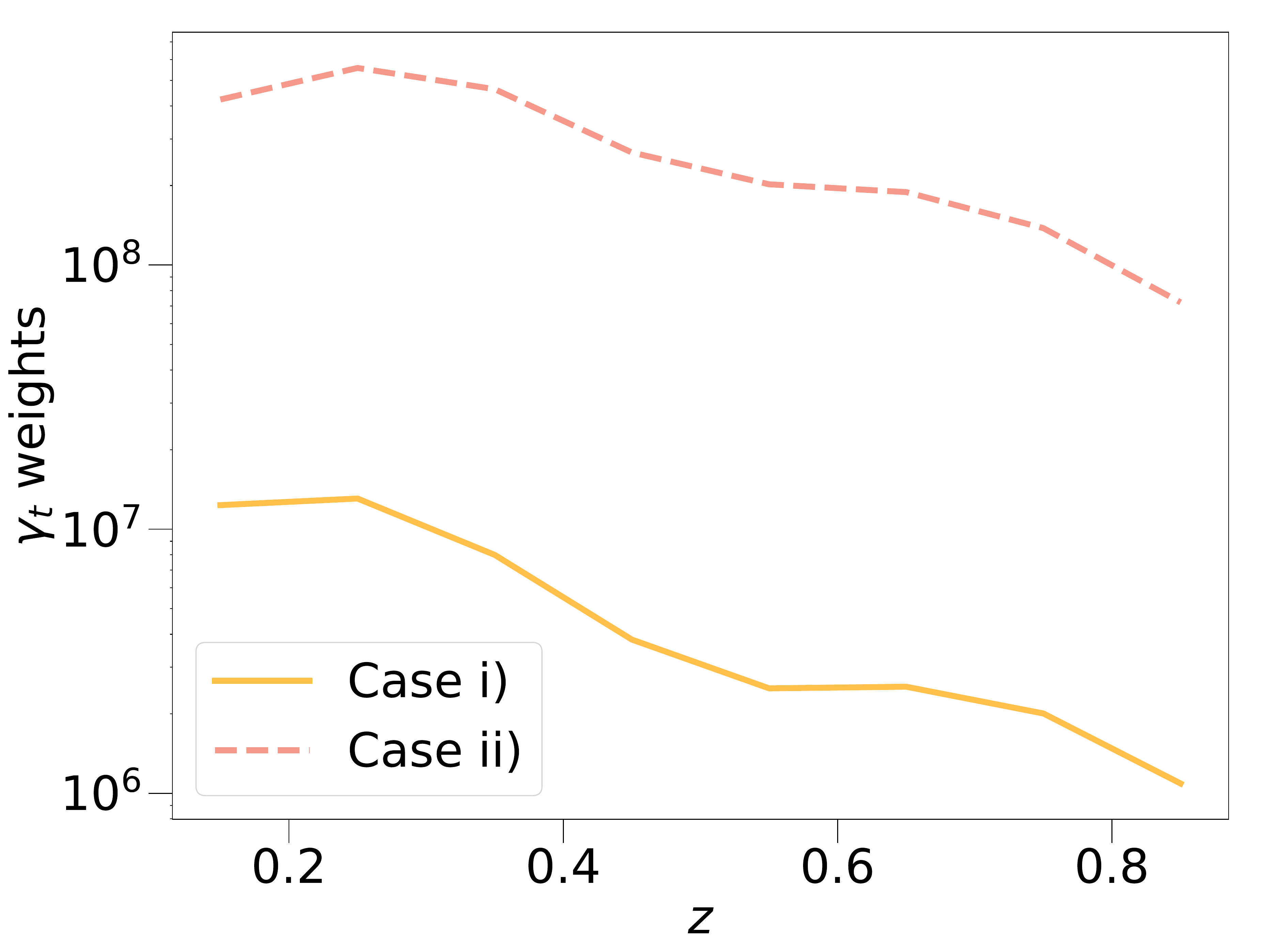}
    \includegraphics[width= \columnwidth]{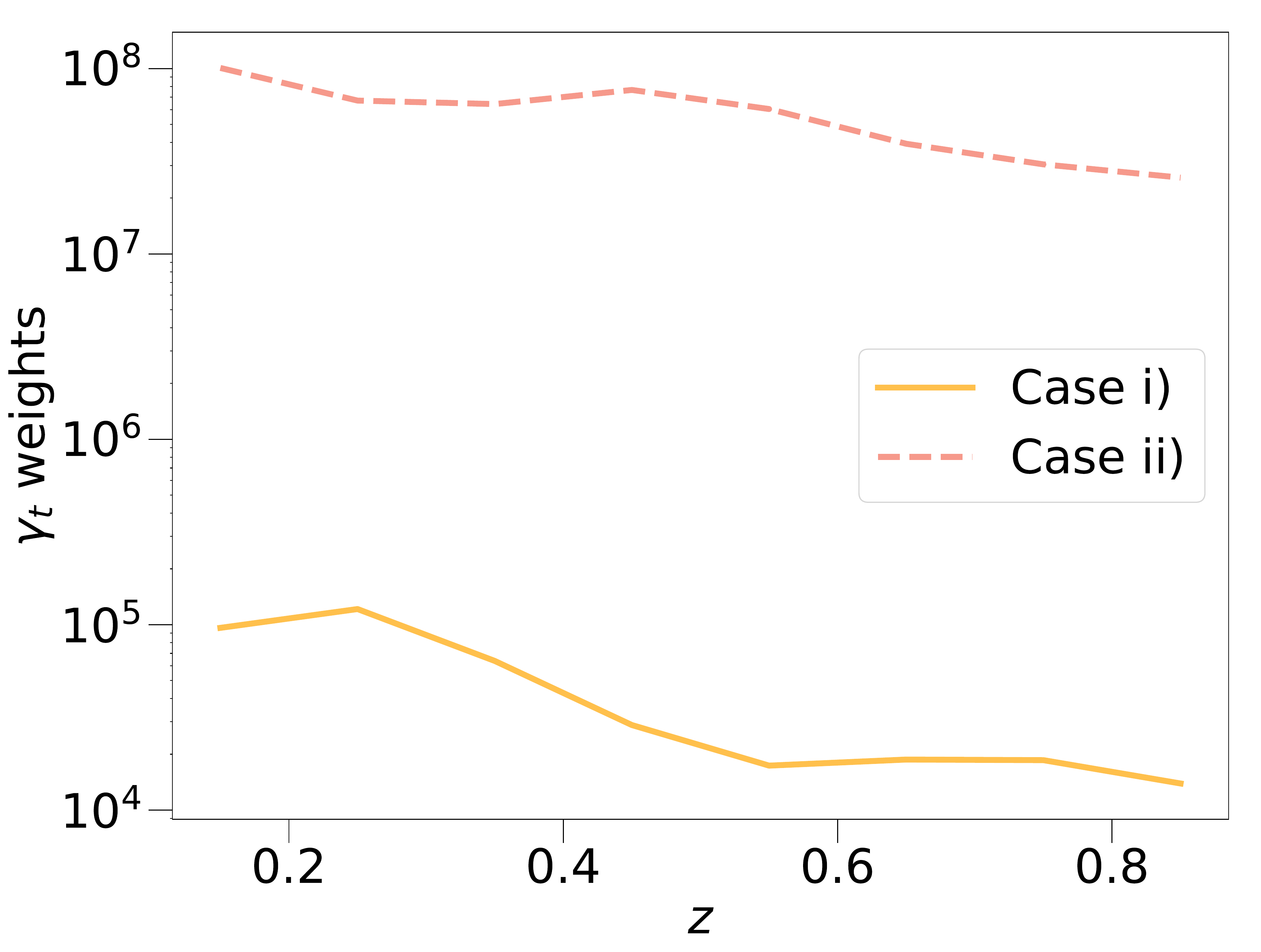}
    \caption{Comparison of the $\gamma_t$ weights as a function of redshift, following  Case i) and Case ii) approaches (solid and dashed lines), using  $\theta_\mathrm{eff} = 180^\circ/200$ (top) and  $\theta_\mathrm{eff} = 180^\circ/100$ (bottom) for Case ii), and displaying the weights for $\theta = \theta_{\rm eff}$ for Case i).  Only the components of the weight matrix for $\gamma_t$, and with respect to $\sigma_8$ are displayed here. }
    \label{fig:weights_gt}
\end{figure}

\begin{figure}
    \centering
     \includegraphics[width= \columnwidth]{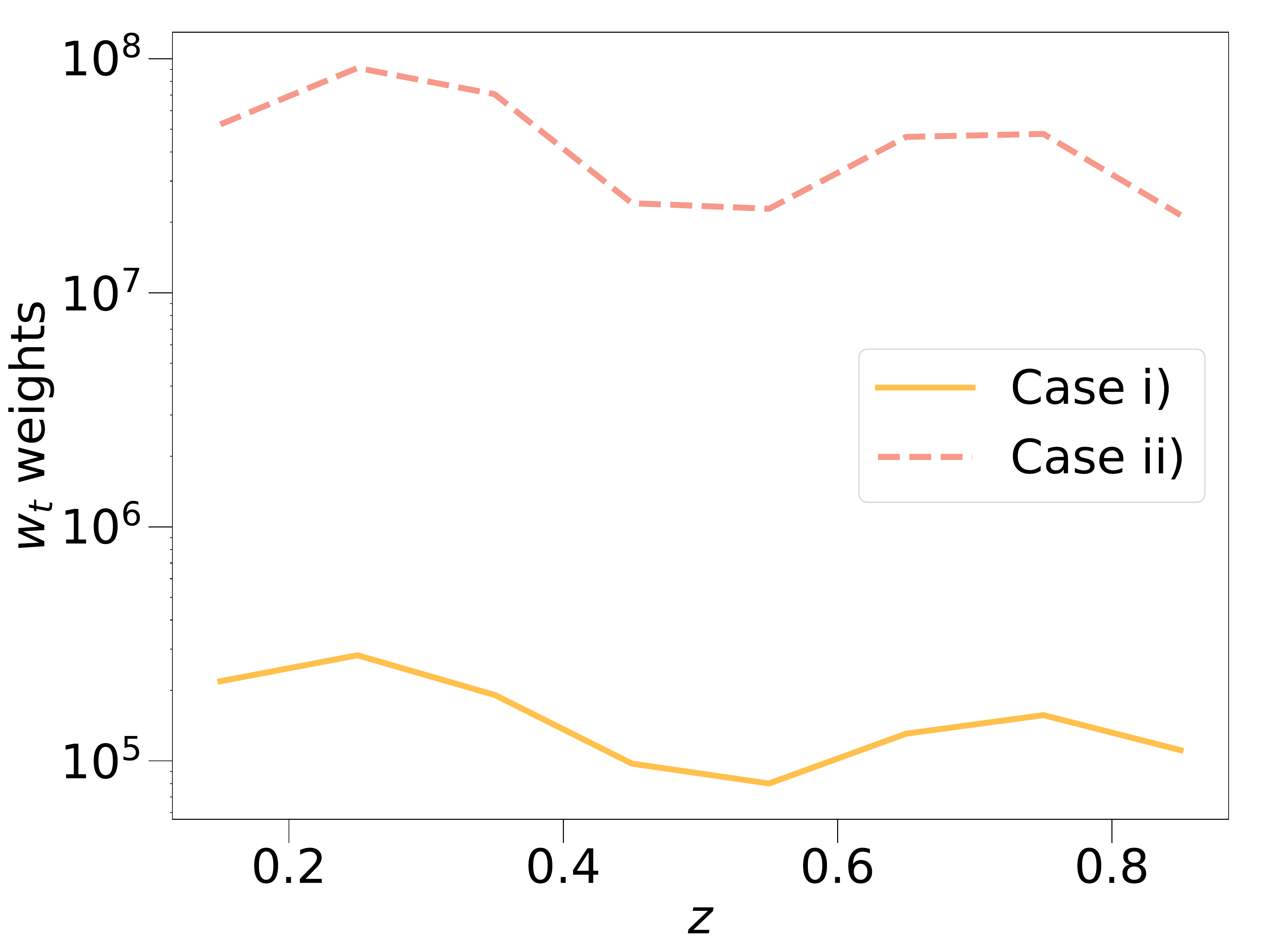}
    \includegraphics[width= \columnwidth]{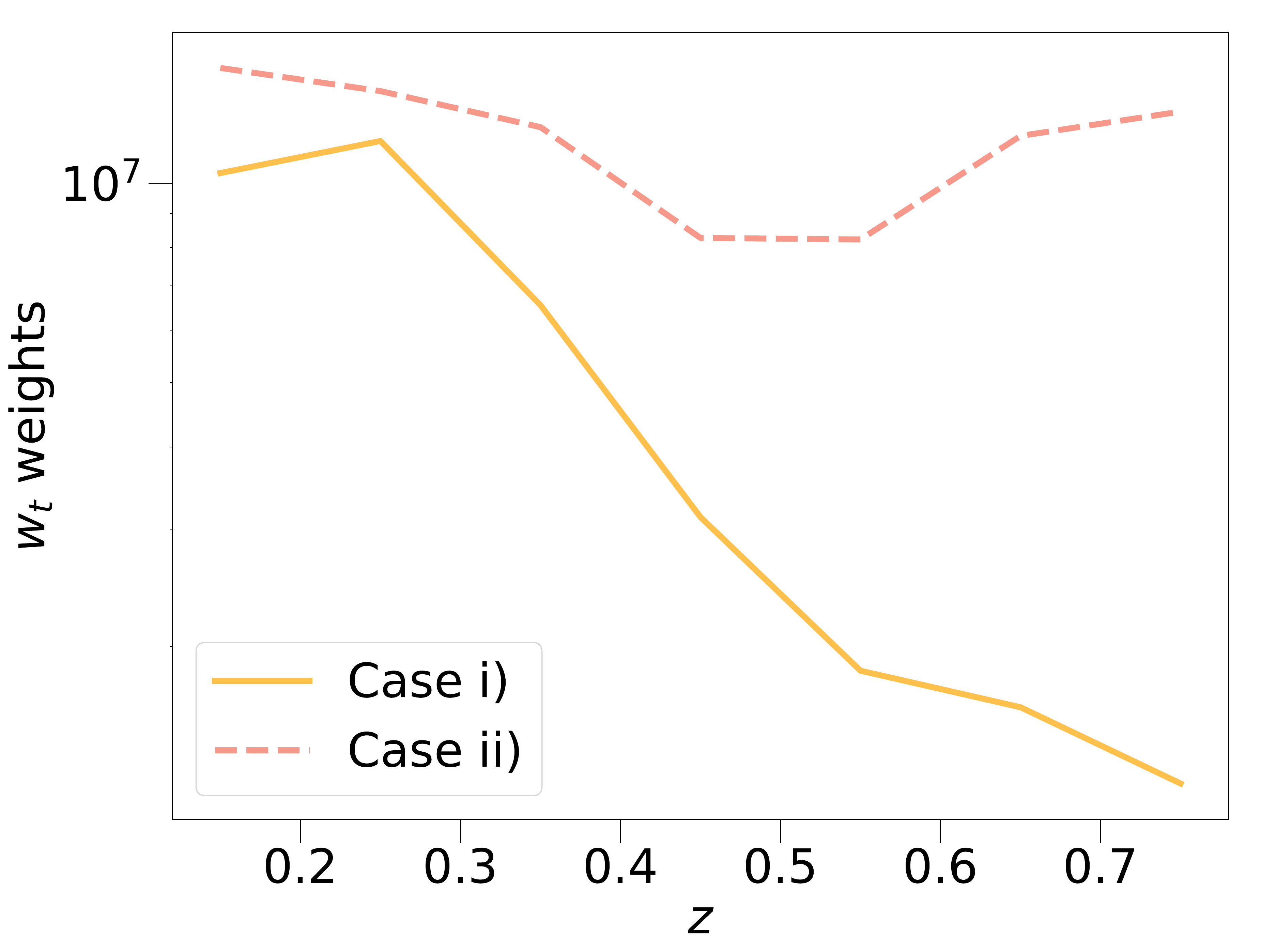}
    \caption{Comparison of the $\omega(\theta)$ weights as a function of redshift, displayed in the same style as Fig. \ref{fig:weights_gt}.}
    \label{fig:weights_w}
\end{figure}

We computed the set of weight matrices for these four parameters, $\mathbf{W}$ and $\mathbf{W}_\mathrm{app}$, as derived for Case i) and Case ii).  In Figs. \ref{fig:weights_gt} and \ref{fig:weights_w} we compare the weights of Case i) (solid lines) and Case ii) (dashed lines).
Fig \ref{fig:weights_gt} displays the component of the weight matrix derived from $\gamma_t$, while Fig. \ref{fig:weights_w} shows the weights corresponding to $\omega$. In both Figures we display only the component of the weights related to $\sigma_8$, noting that similar forms are obtained for the bias component.

We select two different effective scales corresponding to angular wavenumbers $\ell = (100, 200)$ to compare the weights of Case i) and Case ii), which are representative of the range of scales we are fitting, analogous to the \cite{1994ApJ...426...23F} scheme. The Case ii) weights are obtained fixing an effective scale $\theta_{\rm eff} = 180^\circ/\ell$ in the computation of the weight matrix. Case i) weights are computed as a function of scale, for comparison we plot the Case i) weights for $\theta_\mathrm{eff} = 180^\circ/200$ and $\theta_\mathrm{eff} = 180^\circ/100$ components.

\rossi{For the purposes of these comparisons,  we are plotting the weight functions relative to a particular scale $\theta_\mathrm{eff}$, for both Case i) and ii). However, those weight functions are not equivalent, as the scale dependent weights in Case i) are derived using the full covariance across all scales, while the Case ii) weights only consider the value of the variance at a single scale. In practice, while applying Case i) weights to the data, part of the information about $\theta_\mathrm{eff}$ is contained in different weighted modes, given that the covariance across all scales is non-diagonal, while Case ii) assume the same weight function across all scales.  Further, the shape of the weights with redshift depends on the balance of sample variance and noise, which depends on scale.  Thus, the choice of fixing a specific scale $\theta_\mathrm{eff}$, or compressing over all scales, impacts the weights.  Although we will show that the eventual cosmological result is not sensitive to these differences.}

In general, the trend of the weights with redshift is determined by how the covariance component interacts with the derivative of the models.
As expected, we find that the $\gamma_t$ weights shown in Fig. \ref{fig:weights_gt} decrease with redshift: following the trend of the galaxy-galaxy lensing signal to decrease with redshift, given that fewer sources lie behind higher-redshift lenses.  For $\theta_\mathrm{eff} = 180^\circ/100$ we notice a slightly different trend with redshift for $\omega $ (bottom panel of Figure \ref{fig:weights_w}).

We do not find a significant difference in the trend of the weights with redshift for different cases, noting that the normalization of the weights is free\footnote{The normalization is free for the total weight function, applied to $\gamma_t + \omega $; Fig. \ref{fig:weights_gt}, \ref{fig:weights_w} show individual component of the weights ($\gamma_t, \omega$), for which the normalization is not arbitrary} in the derivation.  We can say that the trend of the weights does not vary significantly as a function of scale.

%\cab{(Add a comment on Fig.5 as well.) }\rossi{see above as well, while adding comment on wtheta}

These weights can be applied to obtain the individual galaxy weights by following the procedure described in Sec. \ref{subsec:weigal}.  Given that these treatments are equivalent, we just present the correlation function fits in this section.


\subsection{Fits to compressed correlations}
\label{subsec:compfit}
 
We applied the weight matrices $\mathbf{W}$ and $\mathbf{W}_\mathrm{app}$ to the uncompressed data vectors $\mathbf{X}$ and $\mathbf{M}$ (Eq. \ref{eq:weightscale} and Eq. \ref{eq:weightapprox}) to obtain the compressed datasets $\mathbf{Y}$,  $\mathbf{Y}_\mathrm{app}$ and compressed models $\Modcomp, \Modcomp_\mathrm{app}$.  Given the $R(\theta) > 5 \, h^{-1} \mathrm{Mpc}$ scale cut, only data from a subset of redshift bins in $\mathbf{X}$ contributes to the corresponding scale bin of the compressed data $\mathbf{Y}$, consistently with the uncompressed case. 
The length of the compressed data vector is $2\times 153$, 
the $\theta$ ranges considered for each redshift bin are displayed in Table \ref{table:tablethetarange}. 
\begin{table}
    \centering
    \begin{tabular}{c|c}
  \hline
  $z$ range  & $\theta$ range [deg]   \\
  \hline
  $0.1< z <0.2$       & $0.67 < \theta < 2.67 $\\
  $0.2< z <0.3$       & $0.42 < \theta < 2.67 $\\
  $0.3< z <0.4$       & $0.34 < \theta < 2.67 $\\
  $0.4< z <0.5$       & $0.27 < \theta < 2.67 $\\
  $0.5< z <0.6$       & $0.21 < \theta < 2.67 $\\
  $0.6< z <0.7$       & $0.21 < \theta < 2.67 $\\
  $0.7< z <0.8$       & $0.16 < \theta < 2.67 $\\
  $0.8< z <0.9$       & $0.16 < \theta < 2.67 $\\
  \hline
    \end{tabular}
    \caption{Angular scales considered for $R(\theta) > 5 h^{-1}$ Mpc for each redshift bin, displayed using the range of central angular separation bin values included in our analysis.}
    \label{table:tablethetarange}
\end{table}

\begin{figure}
    \centering
    \includegraphics[width= \columnwidth]{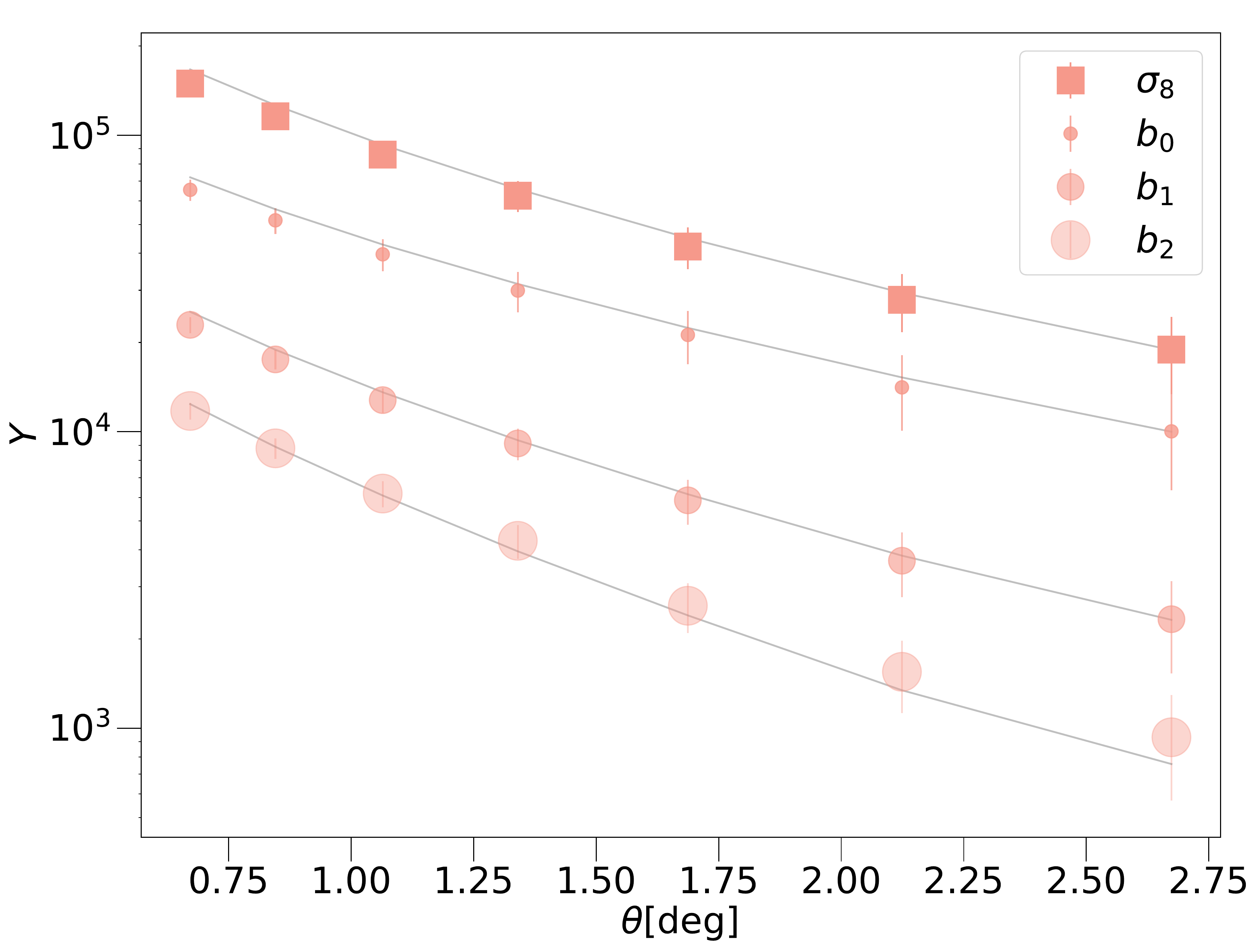}
    \caption{The compressed correlation functions $Y(\theta)$ weighted with respect to the different parameters ($\sigma_8, b_0, b_1, b_2$) for a scale cut $R(\theta) > 5 h^{-1}$ Mpc. The best-fit model is plotted in grey. The error bars are computed from the diagonal of the compressed covariance matrix.}
    \label{fig:bestfits2}
\end{figure}

Figure \ref{fig:bestfits2} displays the compressed datasets $\textbf{Y}$, obtained using Eq. \ref{eq:ycompress}, with respect to the different parameters of interest ($\sigma_8, b_0, b_1, b_2$; red squares and circles). The error bars are computed from the diagonal of the compressed covariance matrix.

\begin{figure}
    \centering
    \includegraphics[width= \columnwidth]{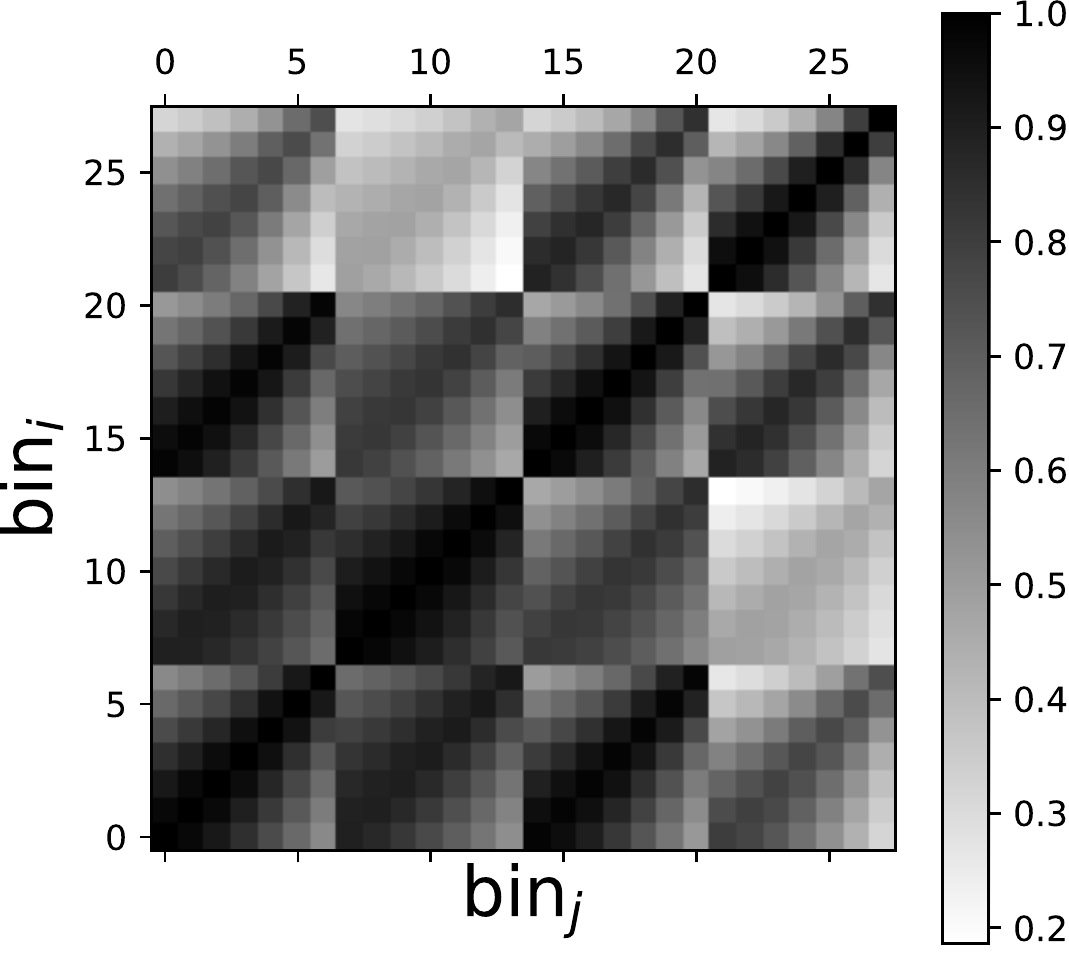}
    \caption{The correlation matrix of the compressed correlation functions for  $N_p = 4$  free parameters. The bins are ordered in scale and in the four different parameters: $i$ and $j$ range from 1 to $N_\theta \times N_p$ as illustrated in Eq. \ref{eq:comprcov}.}
    \label{fig:covariance_compressed}
\end{figure}

We computed the compressed covariances for Cases i) and ii) following Eq. \ref{eq:comprcov} and Eq. \ref{eq:apprcomprcov}. Fig. \ref{fig:covariance_compressed} displays the covariance for the compressed data sample, as correlation matrix. For simplicity, we show all the scales $N_\theta$ before the scale cut, which depends on redshift. For both Case i) and ii) the $\chi^2$ statistic is built as, 
\begin{equation}\label{eq:comprchi2}
\begin{split}
        \chi^2 =  \mathbf{(\Modcomp_{ Y}- Y)^\mathrm{T} C_{ Y}^{-1} (\Modcomp_{ Y}-Y) } ,
\end{split}
\end{equation}
where for Case ii) we use $\textbf{Y}_{\rm app}$ instead of $\textbf{Y}$.

We fit for $\sigma_8$ and the quadratic bias parameters, finding mean results $\sigma_8 = \rm{0.760 } \pm \rm{0.0468}$ and $\chi^2/{\rm d.o.f.} = 0.88$  for Case i), and $\sigma_8 = \rm{0.764 } \pm \rm{0.0485}$ and $\chi^2/{\rm d.o.f.} = 0.82$ for Case ii) across the 18 realizations.  We do not find a significant difference in the error in $\sigma_8$ for Case i) with respect to Case ii), meaning that the weights do not lose their optimality when an effective scale is fixed, similar to the FKP weights.

\rossi{In Figure \ref{fig:bias_quad} we compare the results for the bias fits, using the methods described above. We do not include Case ii), as the results are identical to Case i). We fit for a linear, quadratic and cubic evolution for $b(z)$ (grey and yellow and pink shades) using  Case i) compression, and we compare  them with the results from the uncompressed analysis (black dots).  The confidence regions are computed by reconstructing $b(z)$ from the fit of $b_0, b_1, b_2$ in Eq. \ref{eq:bias}, taking the standard deviation of $b(z)$ at each redshift.} We find that the quadratic $b(z)$ gives an appropriate description of the bias-redshift relation $b(z)$, consistent with the uncompressed analysis.  Using a linear evolution for $b(z)$ results in a larger error on $\sigma_8$, probably due to the fact that a linear evolution model is not describing the bias-redshift relation sufficiently accurately,
$\sigma_8= 0.787 \pm 0.053 $ and $\chi^2/{\rm d.o.f.} = 0.82$.  We tested a third-order bias-redshift relation in Case v), finding no significant difference in the constraints with respect Case i).  
 
Fig. \ref{fig:bestfits2} shows the compressed statistics $\textbf{Y}$ for Case i) (red circles), compared to the best-fit model (solid grey line), for the range of scales considered.   We plot the different components of $\textbf{Y}$ with respect to $\sigma_8$ and the different bias parameters.  The error bars on the compressed data are plotted as the square root of the diagonal elements of the covariance matrix.  The models provide a good description of the data, reflecting the acceptable value of the $\chi^2$ statistic.

\subsection{Comparison between analyses}
\label{subsec:compare}
\begin{figure}
    \centering
    \includegraphics[width= \columnwidth]{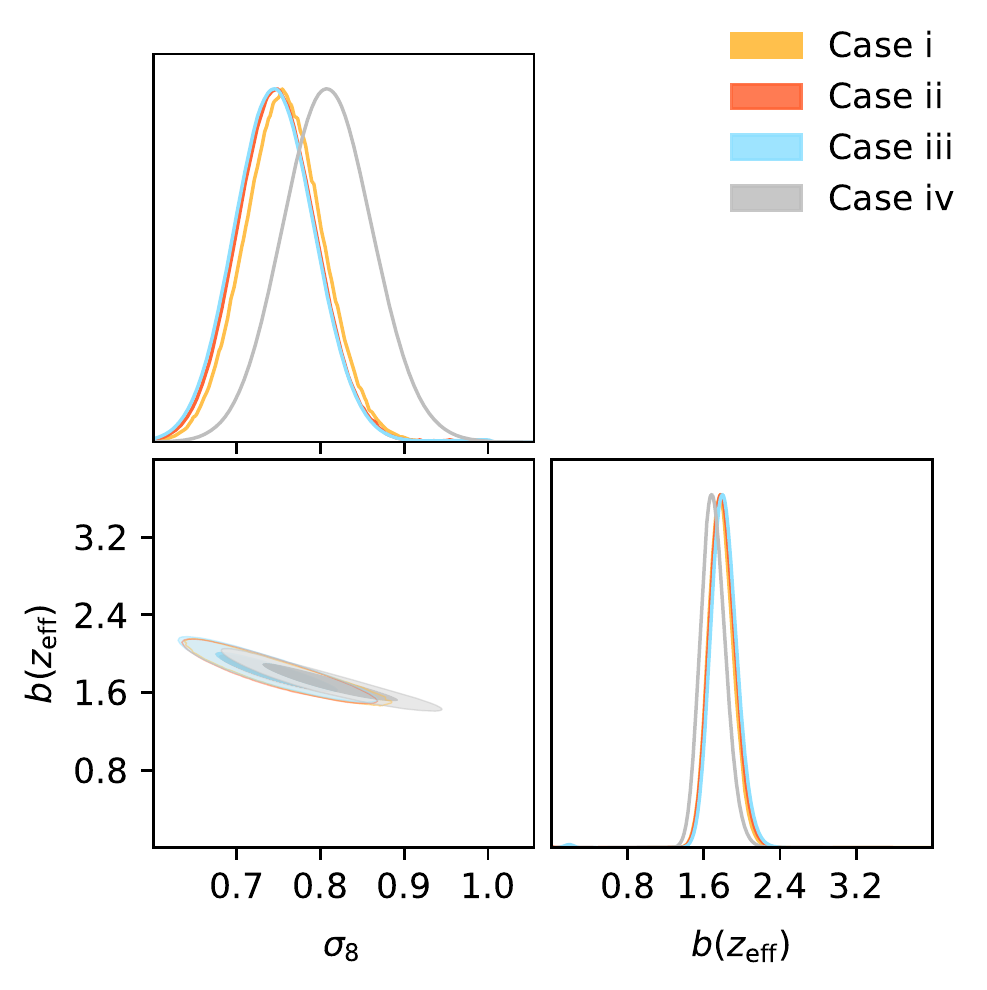}
    \caption{\rossi{Likelihood contours for $\sigma_8$ and bias from the different methods: Case i-ii-iii-iv. 
    We present the results obtained for one of the region considered. 
    For the bias, for Case i-ii-iii we compare $b(z_\mathrm{eff}) = b_0 + b_1 z_\mathrm{eff} + b_2 z^2_\mathrm{eff}$ with $z_\mathrm{eff} = 0.5$, and  $b(z_\mathrm{eff}) = b_0 + b_1 z_\mathrm{eff}$ for Case iv, as per Eq. \ref{eq:bias}.  
 }   }
    \label{fig:s8contours}
\end{figure}

\rossi{In Fig \ref{fig:s8contours} we display the likelihood contour for $\sigma_8$ and bias for one of the regions considered, from the different methods considered: Case i-ii-iii-iv (the yellow, red, blue, grey contours respectively).  For the bias, we compare $b(z_\mathrm{eff}) = b_0 + b_1 z_\mathrm{eff} + b_2 z^2_\mathrm{eff}$ with $z_\mathrm{eff} = 0.5$ for for Case i-ii-iii, and  $b(z_\mathrm{eff}) = b_0 + b_1 z_\mathrm{eff}$ for Case iv, as per Eq. \ref{eq:bias}.   }

\begin{figure}
    \centering
    \includegraphics[width= \columnwidth]{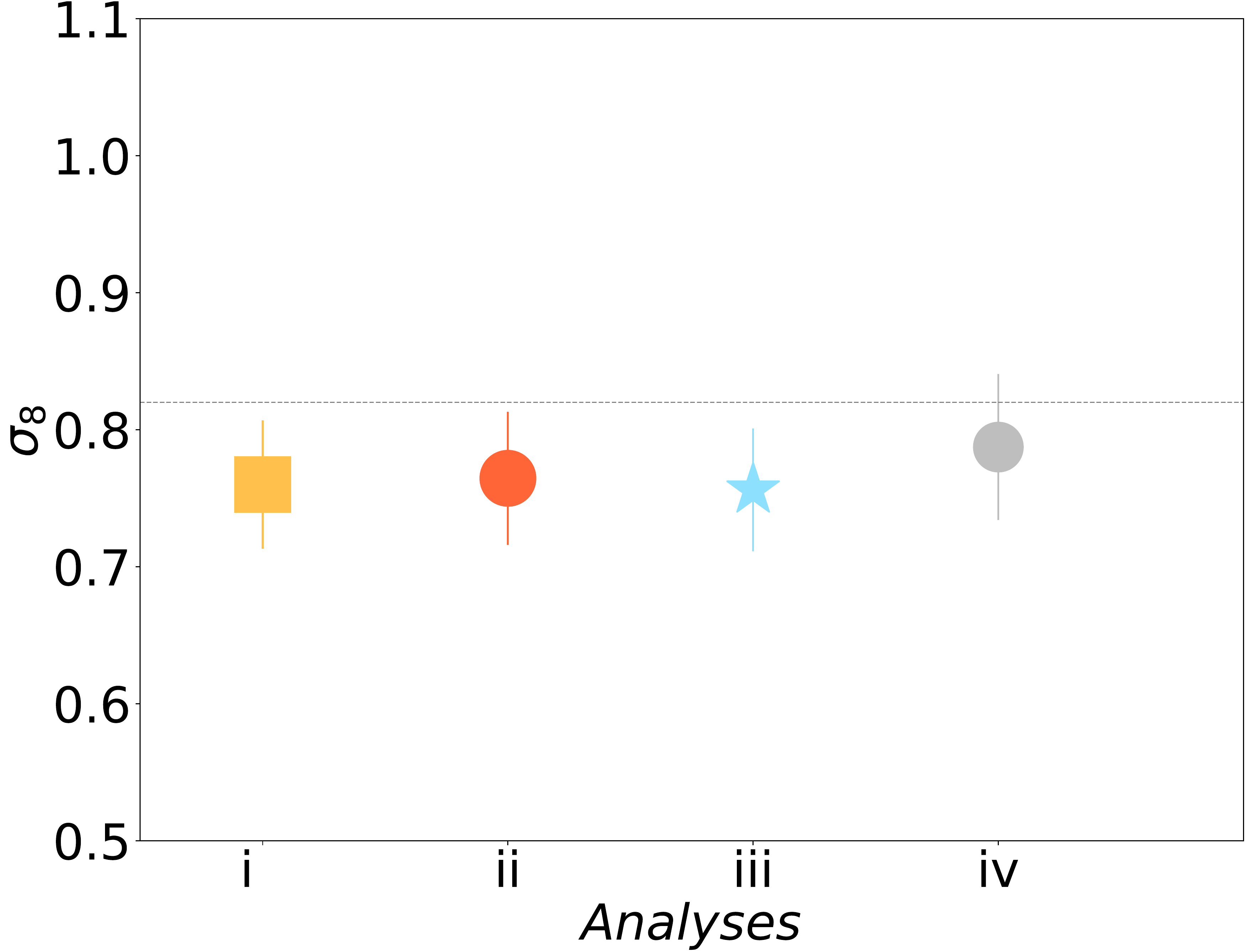}
    \caption{Fitting results for $\sigma_8$ from the different methods, from left to right: Case i-ii-iii-iv. We display  the mean and averaged errors of the 18 regions considered.  The error bars are computed from the standard deviations of the posterior distributions. The values displayed are:  Case i) $\sigma_8 = 0.760 \pm 0.046$, Case ii) $\sigma_8 = 0.764 \pm 0.048$,  Case iii) $\sigma_8 = 0.756 \pm  0.045$. Case iv) $\sigma_8 = 0.787 \pm  0.053$. Dashed grey line indicates the fiducial $\sigma_8$ of the mocks.}
    \label{fig:s8}
\end{figure}

We summarise the results for all cases and regions in Fig. \ref{fig:s8}, where we display the 68\% confidence region for the posterior probability distribution of $\sigma_8$. From left to right, the yellow dot indicates the result from optimal weighting including scale dependence; the red dot indicates the results from a compressed analysis with an effective scale.
The blue point denotes the uncompressed analysis, and the grey point corresponds to a compressed single-scale analysis with a linear redshift evolution for $b(z)$ in the model (as opposed to the fiducial quadratic evolution).  We find consistent results between all types of analysis, and comparable standard deviations for both Cases i,ii and the uncompressed analysis, implying that the compression is almost lossless, even for the single-scale approximation.  We note that an analysis based on the individual galaxy weights, which are derived to match the single-scale approximation, would give the same results.

\begin{figure}
    \centering
    \includegraphics[width=\columnwidth]{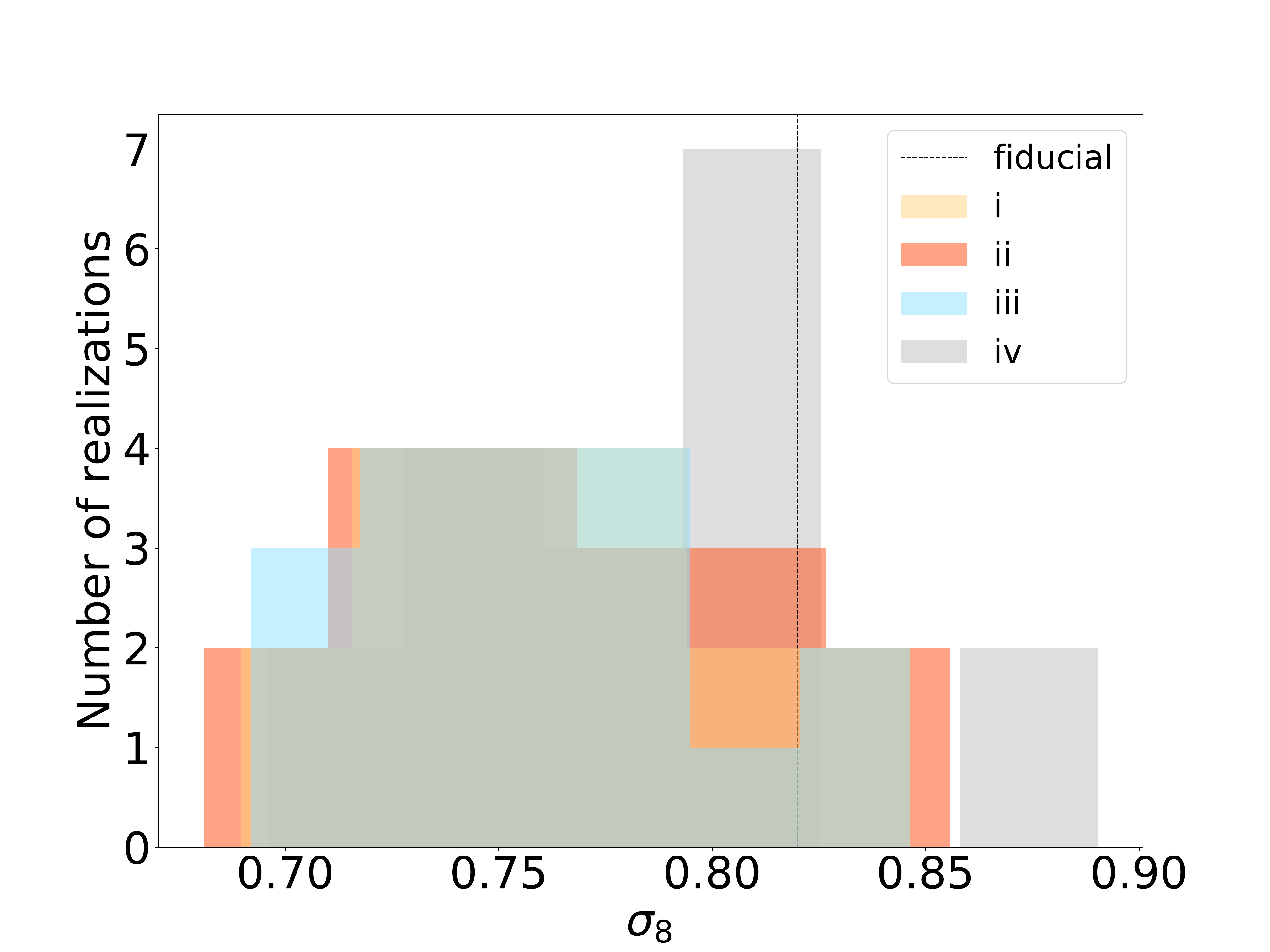}
    \caption{The distribution of the best-fitting $\sigma_8$ across the 18 Buzzard mocks for Case i-ii-iii-iv. Each $\sigma_8$ corresponds to the maximum likelihood value.}
    \label{fig:f_histo}
\end{figure}
\begin{figure}
    \centering
    \includegraphics[width=\columnwidth]{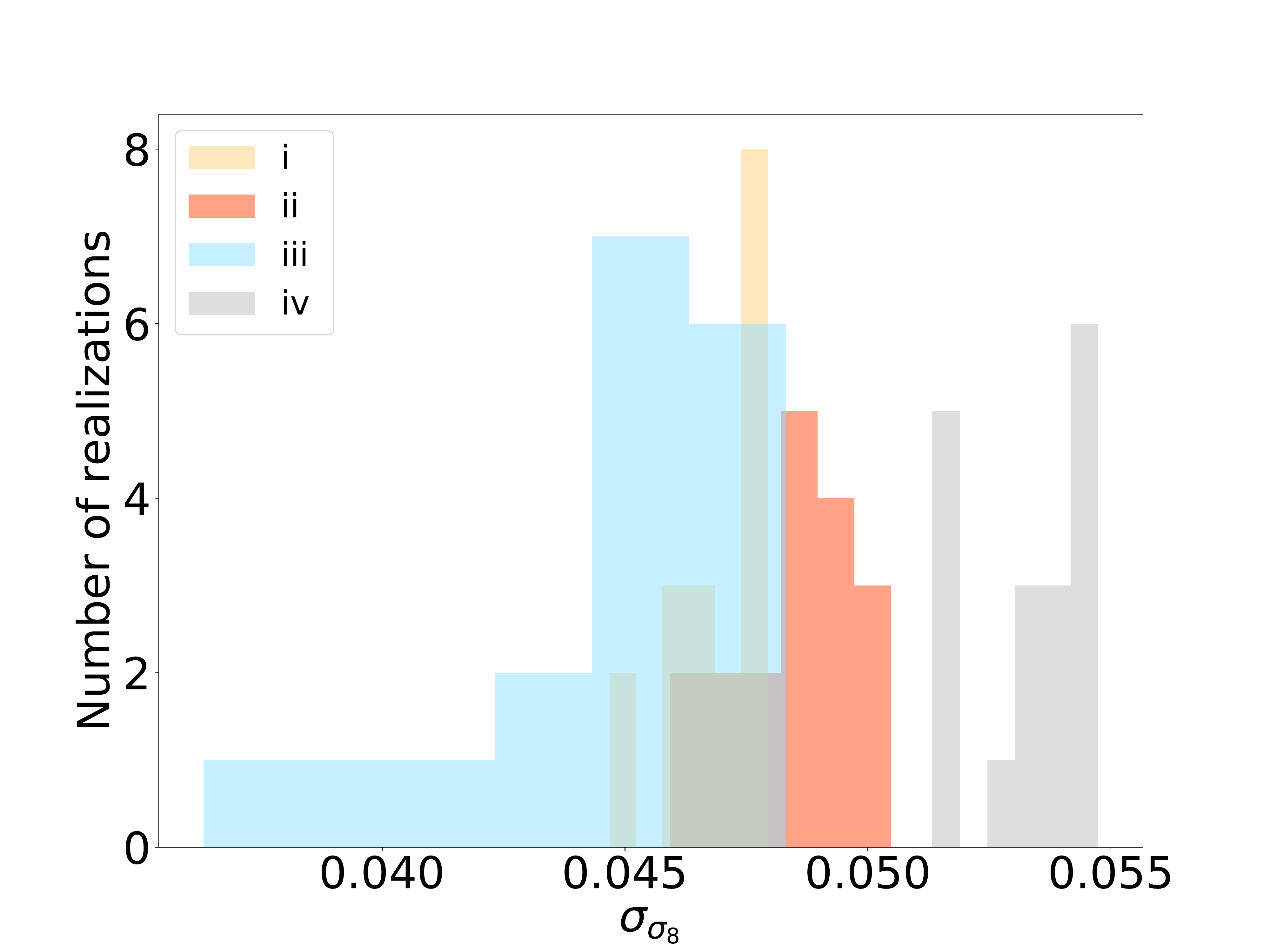}
    \caption{The distribution of the standard deviation of $\sigma_8$ across the 18 Buzzard regions for Case i-ii-iii-iv. Each standard deviation is computed from the posterior distribution of $\sigma_8$.}
    \label{fig:errf_histo}
\end{figure}

\begin{figure}
    \centering
    \includegraphics[width= \columnwidth]{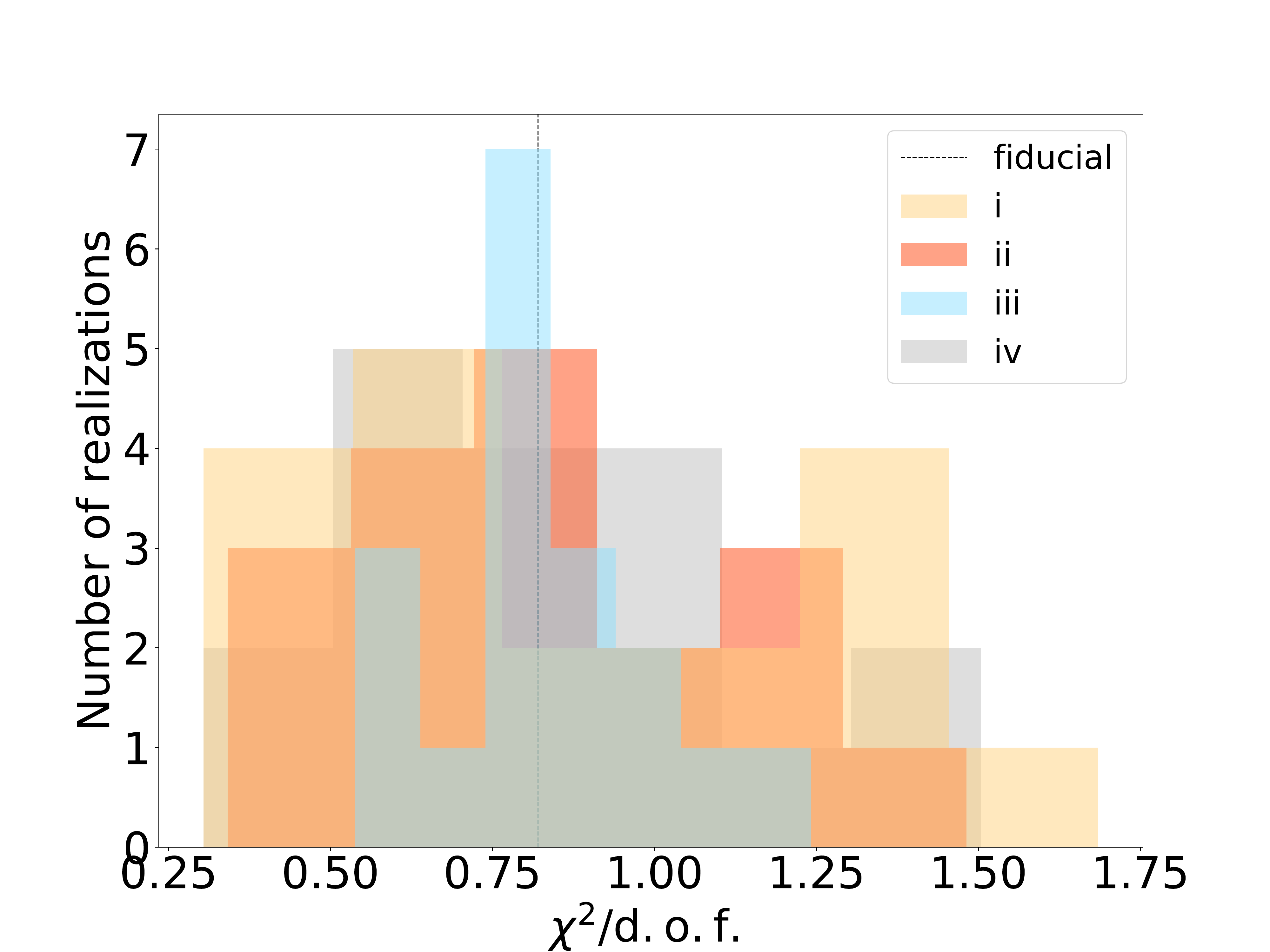}
    \caption{The distribution of reduced $\chi^2$ across the 18 Buzzard regions for Case i-ii-iii-iv.}
    \label{fig:chi_histo}
\end{figure}

The histograms in Fig \ref{fig:f_histo} show the distribution of $\sigma_8$ as measured across the 18 Buzzard regions used in this study, for the four cases analysed. The displayed colours are consistent with Fig. \ref{fig:s8}.  We find consistent values of the mean for all cases.
For robustness check, we repeated Case iii) fits using the same bias redshift relationship as  for Case ii-i), as described in Eq. \ref{eq:bias}, finding identical results on $\sigma_8$ best-fit \rossi{and variance}. We expected this, from Fig. \ref{fig:bias_quad} which shows that a quadratic bias-redshift relationship is well in agreement with the bias values measured across the 8 redshift bins.
Case iv) with linear evolving bias is slightly shifted with respect to the other distributions. As discussed earlier, linear bias evolution is not accurately reproducing the bias parameters fitted in individual redshift bins. We find the best-fitting $\sigma_8$ for all methods to be slightly lower \rossi{(close to 1$\sigma$)} than the fiducial value of the mocks, shown as the black dashed line.  As discussed in Sec. \ref{subsec:uncompfit}, this issue is likely related to the linear bias model adopted in our analysis, and does not affect the conclusions of our current study.

\rossi{Fig. \ref{fig:errf_histo} shows the distribution of the standard deviation for $\sigma_8$ as measured across 18 Buzzard regions. The displayed colours are consistent with Fig. \ref{fig:s8}.  The compressed analysis (both optimal and single-scale, yellow and red histograms) gives similar results to the uncompressed analysis (blue histogram):
 Case i) $\sigma_8 = 0.760 \pm 0.046$, Case ii) $\sigma_8 = 0.764 \pm 0.048$  and Case iii) $\sigma_8 = 0.756 \pm  0.045$.
The analysis with a linear redshift evolution for $b(z)$ (grey histogram) leads to higher values of standard deviation: Case iv) $\sigma_8 = 0.787 \pm  0.053$. }

Fig \ref{fig:chi_histo} displays the distribution of $\chi^2$ across all the regions (with the same colour as for Fig. \ref{fig:f_histo}). 
We find that all realizations are consistent with reduced $\chi^2 = 1$, noticing a slightly higher mean $\chi^2$ for the linear-evolving bias case.
Case iii) shows a narrower distribution of $\chi^2/\mathrm{d.o.f,}$ with respect the other Cases. This is because Case iii) has a higher number of free parameters (9), and the variance of the $\chi^2$ scales with the inverse of $\mathrm{d.o.f.}$

\section{Conclusion}
\label{sec:conc}

In this work we used simulated datasets of future combined galaxy clustering and weak lensing surveys, representing the region of overlap between DESI Y1 and imaging surveys such as DES, HSC and KiDS, to explore data compression and optimal weighting algorithms for extracting cosmological parameters from these combined probes.  We derived data-compression schemes which can be applied to galaxy-galaxy lensing and clustering statistics, extending the work of \cite{2020MNRAS.498.2948R} into configuration space.  For this test case we focussed on a small parameter set consisting of $\sigma_8$ and the galaxy bias parameters (whose degeneracy can be broken by combined probes), noting that these techniques can be readily extended to larger parameter sets.

We investigated different types of compression, in particular comparing a sub-optimal approach where an effective angular scale is selected to derive the weights, with an optimal scheme where the scale dependence is included in the weights, finding no significant difference in the results \rossi{(for scale dependent weights: $\sigma_8 = 0.760 \pm 0.046$, for effective angular scale weights: $\sigma_8 = 0.764 \pm 0.048$).  This allows us to use the single-scale weights to derive a weighting scheme to be applied to individual galaxies rather than correlation functions, producing near-optimal results in this case. Both compression approaches (single-scale and optimal) lead to a standard deviation of $\sigma_8$ very similar to the uncompressed analysis ($\sigma_8 = 0.756 \pm  0.045$). This means that in both cases our compression is preserving the information relevant to our scientific goal.}

Future work, as part of the mock-challenge effort to validate DESI Y1 analysis, will involve:
\begin{itemize}
    \item Testing the optimal galaxy redshift weights in the presence of realistic window functions,
    \item Extending the methodology to more parameters, to allow for wider exploration of the cosmological model,
    \item Extension to more observed statistics, by adding correlation function multipoles which describe redshift-space distortions,
    \item Testing different fitting ranges and non-linear models for the galaxy bias, to get unbiased constraints of $\sigma_8$ and other parameters.
    \item \rossi{Including cosmic shear to our analysis, performing a full 3x2 point correlation function analysis.}
    \item \rossi{As DESI has multiple targets observed, a potential extension of this work may apply the data-compression methodology to multi-tracer analyses.}
\end{itemize}
This program of work will calibrate data-compression techniques for creating novel tests of the cosmological model using combined probes.

\section*{Acknowledgements}
This research was partially funded by the Australian Government through Australian Research Council Discovery Project DP160102705.  Further, this research was supported partially by the Australian Government through the Australian Research Council's Australian Research Council Laureate Fellowship funding scheme (project FL180100168). 
\rossi{This research is supported by the Director, Office of Science, Office of High Energy Physics of the U.S. Department
of Energy under Contract No. DE–AC02–05CH11231, and by the National Energy Research Scientific Computing
Center, a DOE Office of Science User Facility under the same contract; additional support for DESI is provided by
the U.S. National Science Foundation, Division of Astronomical Sciences under Contract No. AST-0950945 to the
NSF’s National Optical-Infrared Astronomy Research Laboratory; the Science and Technologies Facilities Council of
the United Kingdom; the Gordon and Betty Moore Foundation; the Heising-Simons Foundation; the French Alternative Energies and Atomic Energy Commission (CEA); the National Council of Science and Technology of Mexico
(CONACYT); the Ministry of Science and Innovation of Spain (MICINN), and by the DESI Member Institutions:
https://www.desi.lbl.gov/collaborating-institutions.
The authors are honored to be permitted to conduct scientific research on Iolkam Du’ag (Kitt Peak), a mountain
with particular significance to the Tohono O’odham Nation.}
%%%%%%%%%%%%%%%%%%%%%%%%%%%%%%%%%%%%%%%%%%%%%%%%%%
\section*{Data Availability}

The data underlying this article will be shared on reasonable request to the corresponding author.


%  These Macros are taken from the AAS TeX macro package version 4.0.
%  Include this file in your LaTeX source only if you are not using
%  the AAS TeX macro package and need to resolve the macro definitions
%  in the BibTeX entries returned by the ADS abstract service.
%
%  For more information on the AASTeX macro package, please see the URL
%	http://www.aas.org/publications/aastex.html
%  For more information about ADS abstract server, please see the URL
%	http://adswww.harvard.edu/ads_abstracts.html
%

% Abbreviations for journals.  The object here is to provide authors
% with convenient shorthands for the most "popular" (often-cited)
% journals; the author can use these markup tags without being concerned
% about the exact form of the journal abbreviation, or its formatting.
% It is up to the keeper of the macros to make sure the macros expand
% to the proper text.  If macro package writers agree to all use the
% same TeX command name, authors only have to remember one thing, and
% the style file will take care of editorial preferences.  This also
% applies when a single journal decides to revamp its abbreviating
% scheme, as happened with the ApJ (Abt 1991).

\def\jnl@style{\it}
%commente par Seb
\def\aaref@jnl#1{{\jnl@style#1}}
%ref remplace par aaref pour eviter conflit...

\def\aaref@jnl#1{{\jnl@style#1}}

\def\aj{\aaref@jnl{AJ}}                   % Astronomical Journal
\def\araa{\aaref@jnl{ARA\&A}}             % Annual Review of Astron and Astrophys
\def\apj{\aaref@jnl{ApJ}}                 % Astrophysical Journal
\def\apjl{\aaref@jnl{ApJ}}                % Astrophysical Journal, Letters
\def\apjs{\aaref@jnl{ApJS}}               % Astrophysical Journal, Supplement
\def\ao{\aaref@jnl{Appl.~Opt.}}           % Applied Optics
\def\apss{\aaref@jnl{Ap\&SS}}             % Astrophysics and Space Science
\def\aap{\aaref@jnl{A\&A}}                % Astronomy and Astrophysics
\def\aapr{\aaref@jnl{A\&A~Rev.}}          % Astronomy and Astrophysics Reviews
\def\aaps{\aaref@jnl{A\&AS}}              % Astronomy and Astrophysics, Supplement
\def\azh{\aaref@jnl{AZh}}                 % Astronomicheskii Zhurnal
\def\baas{\aaref@jnl{BAAS}}               % Bulletin of the AAS
\def\jrasc{\aaref@jnl{JRASC}}             % Journal of the RAS of Canada
\def\memras{\aaref@jnl{MmRAS}}            % Memoirs of the RAS
\def\mnras{\aaref@jnl{MNRAS}}             % Monthly Notices of the RAS
\def\pra{\aaref@jnl{Phys.~Rev.~A}}        % Physical Review A: General Physics
\def\prb{\aaref@jnl{Phys.~Rev.~B}}        % Physical Review B: Solid State
\def\prc{\aaref@jnl{Phys.~Rev.~C}}        % Physical Review C
\def\prd{\aaref@jnl{Phys.~Rev.~D}}        % Physical Review D
\def\pre{\aaref@jnl{Phys.~Rev.~E}}        % Physical Review E
\def\prl{\aaref@jnl{Phys.~Rev.~Lett.}}    % Physical Review Letters
\def\pasp{\aaref@jnl{PASP}}               % Publications of the ASP
\def\pasj{\aaref@jnl{PASJ}}               % Publications of the ASJ
\def\qjras{\aaref@jnl{QJRAS}}             % Quarterly Journal of the RAS
\def\skytel{\aaref@jnl{S\&T}}             % Sky and Telescope
\def\solphys{\aaref@jnl{Sol.~Phys.}}      % Solar Physics
\def\sovast{\aaref@jnl{Soviet~Ast.}}      % Soviet Astronomy
\def\ssr{\aaref@jnl{Space~Sci.~Rev.}}     % Space Science Reviews
\def\zap{\aaref@jnl{ZAp}}                 % Zeitschrift fuer Astrophysik
\def\nat{\aaref@jnl{Nature}}              % Nature
\def\iaucirc{\aaref@jnl{IAU~Circ.}}       % IAU Cirulars
\def\aplett{\aaref@jnl{Astrophys.~Lett.}} % Astrophysics Letters
\def\apspr{\aaref@jnl{Astrophys.~Space~Phys.~Res.}}
                % Astrophysics Space Physics Research
\def\bain{\aaref@jnl{Bull.~Astron.~Inst.~Netherlands}} 
                % Bulletin Astronomical Institute of the Netherlands
\def\fcp{\aaref@jnl{Fund.~Cosmic~Phys.}}  % Fundamental Cosmic Physics
\def\gca{\aaref@jnl{Geochim.~Cosmochim.~Acta}}   % Geochimica Cosmochimica Acta
\def\grl{\aaref@jnl{Geophys.~Res.~Lett.}} % Geophysics Research Letters
\def\jcp{\aaref@jnl{J.~Chem.~Phys.}}      % Journal of Chemical Physics
\def\jgr{\aaref@jnl{J.~Geophys.~Res.}}    % Journal of Geophysics Research
\def\jqsrt{\aaref@jnl{J.~Quant.~Spec.~Radiat.~Transf.}}
                % Journal of Quantitiative Spectroscopy and Radiative Transfer
\def\memsai{\aaref@jnl{Mem.~Soc.~Astron.~Italiana}}
                % Mem. Societa Astronomica Italiana
\def\nphysa{\aaref@jnl{Nucl.~Phys.~A}}   % Nuclear Physics A
\def\physrep{\aaref@jnl{Phys.~Rep.}}   % Physics Reports
\def\physscr{\aaref@jnl{Phys.~Scr}}   % Physica Scripta
\def\planss{\aaref@jnl{Planet.~Space~Sci.}}   % Planetary Space Science
\def\procspie{\aaref@jnl{Proc.~SPIE}}   % Proceedings of the SPIE
\def\jcap{\aaref@jnl{J. Cosmology Astropart. Phys.}}
                % Journal of Cosmology and Astroparticle Physics

\let\astap=\aap
\let\apjlett=\apjl
\let\apjsupp=\apjs
\let\applopt=\ao

\newcommand{\mpc}{\, {\rm Mpc}}
\newcommand{\hmpc}{\, h^{-1} \mpc}
\newcommand{\ihmpc}{\, h\, {\rm Mpc}^{-1}}
\newcommand{\ikms}{\, {\rm s\, km}^{-1}}
\newcommand{\kms}{\, {\rm km\, s}^{-1}}
\newcommand{\hkpc}{\, h^{-1} \kpc}
\newcommand{\lya}{Ly$\alpha$\ }
\newcommand{\lyb}{Lyman-$\beta$\ }
\newcommand{\lyaf}{Ly$\alpha$ forest}
\newcommand{\lr}{\lambda_{{\rm rest}}}
\newcommand{\bF}{\bar{F}}
\newcommand{\bS}{\bar{S}}
\newcommand{\bC}{\bar{C}}
\newcommand{\bB}{\bar{B}}
\newcommand{\vdF}{{\mathbf \delta_F}}
\newcommand{\vdS}{{\mathbf \delta_S}}
\newcommand{\vdf}{{\mathbf \delta_f}}
\newcommand{\vdn}{{\mathbf \delta_n}}
\newcommand{\vdC}{{\mathbf \delta_C}}
\newcommand{\vdX}{{\mathbf \delta_X}}
\newcommand{\xrei}{x_{rei}}
\newcommand{\lrmin}{\lambda_{{\rm rest, min}}}
\newcommand{\lrmax}{\lambda_{{\rm rest, max}}}
\newcommand{\lmin}{\lambda_{{\rm min}}}
\newcommand{\lmax}{\lambda_{{\rm max}}}
\newcommand{\hi}{\mbox{H\,{\scriptsize I}\ }}
\newcommand{\heii}{\mbox{He\,{\scriptsize II}\ }}
\newcommand{\vp}{\mathbf{p}}
\newcommand{\vq}{\mathbf{q}}
\newcommand{\vxperp}{\mathbf{x_\perp}}
\newcommand{\vkperp}{\mathbf{k_\perp}}
\newcommand{\vrperp}{\mathbf{r_\perp}}
\newcommand{\vx}{\mathbf{x}}
\newcommand{\vy}{\mathbf{y}}
\newcommand{\vk}{\mathbf{k}}
\newcommand{\vR}{\mathbf{r}}
\newcommand{\tdtwo}{\tilde{b}_{\delta^2}}
\newcommand{\tstwo}{\tilde{b}_{s^2}}
\newcommand{\tbthree}{\tilde{b}_3}
\newcommand{\tadtwo}{\tilde{a}_{\delta^2}}
\newcommand{\tastwo}{\tilde{a}_{s^2}}
\newcommand{\tabthree}{\tilde{a}_3}
\newcommand{\vnabla}{\mathbf{\nabla}}
\newcommand{\tpsi}{\tilde{\psi}}
\newcommand{\tfnl}{{\tilde{f}_{\rm NL}}}
\newcommand{\gnl}{g_{\rm NL}}
\newcommand{\orderfour}{\mathcal{O}\left(\delta_1^4\right)}
\newcommand{\SDSSPF}{\cite{2006ApJS..163...80M}}
\newcommand{\PF}{$P_F^{\rm 1D}(k_\parallel,z)$}
\newcommand\ionalt[2]{#1$\;${\scriptsize \uppercase\expandafter{\romannumeral #2}}}%  
\newcommand{\vxone}{\mathbf{x_1}}
\newcommand{\vxtwo}{\mathbf{x_2}}
\newcommand{\vRot}{\mathbf{r_{12}}}
\newcommand{\cm}{\, {\rm cm}}

\bibliography{biblio2}

%\begin{thebibliography}{99}
%\end{thebibliography}

%%%%%%%%%%%%%%%%%%%%%%%%%%%%%%%%%%%%%%%%%%%%%%%%%%

%%%%%%%%%%%%%%%%% APPENDICES %%%%%%%%%%%%%%%%%%%%%

\clearpage 
\appendix
\label{sec:appendix}
% \begin{minipage}{2\columnwidth}
% Comparison between results of uncompressed analyses using different scale cuts. $R(\theta) > 5 \, h^{-1} \mathrm{Mpc}$ and $R(\theta) > 5 \, h^{-1} \mathrm{Mpc}$
% \end{minipage}
 % Figure \ref{fig:comparisonscalecuts} shows 
 \begin{figure}\label{fig:comparisonscalecuts}
 \begin{minipage}{2\columnwidth}
    \centering
    \includegraphics[width=\columnwidth]{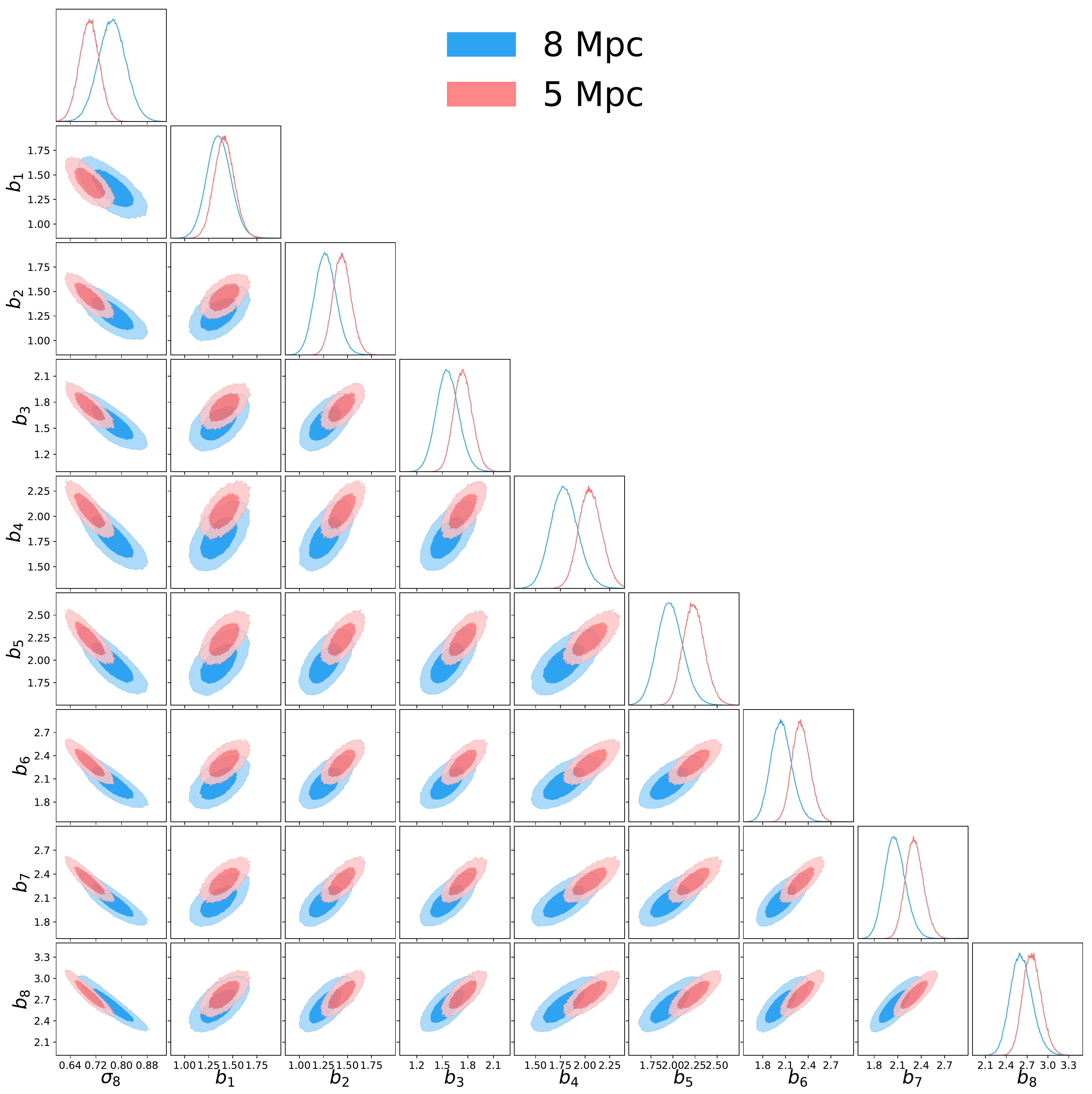}
    \caption{\rossi{Comparison between  the results for the uncompressed analyses using two different scale cuts: $R(\theta) > 5 \, h^{-1} \mathrm{Mpc}$ (red contours) and $R(\theta) > 8 \, h^{-1} \mathrm{Mpc}$ (blue contours). Both analyses assume a galaxy linear bias. As expected the value of $\sigma_8$ recoveconstraints from the $R(\theta) > 8 \, h^{-1} \mathrm{Mpc}$ results is closer to the fiducial cosmology one: $0.82$, as the effect of non-linear galaxy bias is lower than in the $R(\theta) > 5 \, h^{-1} \mathrm{Mpc}$ results. }
    }
    \label{fig:my_label}
 \end{minipage}
\end{figure}

\bsp	% typesetting comment

\label{lastpage}

\end{document}